\documentstyle[11pt,amsmath,amsfonts,amssymb,epsfig]{article}
\def\Pom{I\!\!P}
\begin{document}
\title{$\bar d - \bar u$ asymmetry \\
      and
 semi-inclusive production of pions in DIS}

   \author{A. Szczurek$^{1}$, V. Uleshchenko$^{1,2}$ and J. Speth$^{3}$  \\
   {\it $^{1}$ Institute of Nuclear Physics, PL-31-342 Cracow, Poland  } \\
   {\it $^{2}$ Institute for Nuclear Research, 03-028 Kiev, Ukraine  } \\
   {\it $^{3}$ Institut f\"ur Kernphysik, KFA, J\"ulich, Germany} \\ }

\maketitle

\begin{abstract}
We investigate unambiguities in the extraction of the $\bar d - \bar u$
asymmetry from semi-inclusive production of pions in DIS.
The role of several effects beyond the
quark-parton model (QPM) which lead to
$N_p^{\pi^+} \ne N_n^{\pi^+}$ and
$N_p^{\pi^-} \ne N_n^{\pi^-}$ and may therefore cloud
such an extraction is studied.
The results are discussed in the context of the recent HERMES data.
We find that the interaction of the resolved photon
with the nucleon significantly modifies the observed $\bar d - \bar u$
asymmetry. The exclusive elastic production of $\rho$ mesons
plays a similar role for the large-$z$ data sample.
Our estimate shows a rather small effect of the spectator mechanism.
Nuclear effects in the deuteron also look potentially important
but are difficult to estimate quantitatively.

Throughout the paper we make in addition several general
remarks concerning semi-inclusive and exclusive production of mesons.
\end{abstract}

\vspace{0.5cm}



\section{Introduction}

Since the NMC publication \cite{NMC_GSR} on the Gottfried Sum Rule violation
the effect of $\bar d - \bar u$ asymmetry was in the 90's one
of the most intensively discussed problems of nucleon structure.
The effect is clearly of nonperturbative nature and was
qualitatively explained as due to the
 pion (meson) cloud in the nucleon
(for recent reviews see \cite{MC}). In order to shed more light on
the nature
 of the Gottfried Sum Rule violation two different Drell-Yan
experiments
 were proposed and performed \cite{NA51,E866}.
They measured the ratio
$\sigma_{pd}^{DY} / \sigma_{pp}^{DY}$. The integrated result for
the asymmetry from the more complete Fermilab experiment \cite{E866} is
$\int_0^1 [ \bar d - \bar u ] \; dx$ = 0.09 $\pm$ 0.02, to be
compared with the NMC result:
$\int_0^1 [ \bar d - \bar u ] \; dx$ = 0.148 $\pm$ 0.039.
The NMC asymmetry apears slightly bigger. It was suggested
recently by two of us that the difference can be partly
due to large higher-twist effects for the nonsinglet quantity
$F_2^p - F_2^n$~\cite{SU_part_viol}.

It was proposed in Ref.\cite{LMS91} how to use semi-inclusive
production of pions to extract the asymmetry of light
antiquarks in the nucleon. This method was applied recently by
the HERMES collaboration at HERA \cite{HERMES_dbar_ubar}.

Recent results for semi-inclusive production of pions in polarized 
photoproduction obtained at SLAC \cite{SLAC_photo_semi} have shown that spin 
 asymmetry almost cancels for small transverse momenta of the outgoing 
 pions, which seems to be another nonperturbative effect.  This result was 
interpreted as a large VDM contribution \cite{ACW99} for small transverse 
momenta. Only at large transverse momenta may the perturbative QCD processes 
reveal themselves and only then can one try to disentangle the polarized 
quark distributions in the nucleon.  At low photon virtuality, as in the 
HERMES experiment where
$< \! \! Q^2 \! \!> \: \sim$ 2.3 GeV$^2$, similar nonperturbative
effects can be expected in the unpolarized case.

In the present paper we try to examine the semi-inclusive
production of pions in DIS as a source for measuring
the $\bar d - \bar u$  asymmetry. We investigate several effects,
mostly of nonperturbative nature, which may modify the resulting
asymmetry. In particular, making quantitative estimations,
we focus on conclusions relevant for the HERMES experiment.

\section{Quark-parton model approach}

\subsection{Extraction of the $\bar d - \bar u$ asymmetry}

The most general five-fold cross section for one-particle
semi-inclusive unpolarized
lepton-hadron scattering can be expressed in terms
of four independent semi-inclusive structure functions
(see for instance \cite{LM94}). If the azimuthal correlation
between the lepton scattering plane and the hadron production plane
is not studied, the number of independent structure functions
reduces to two. Then the cross section can be written as
\begin{equation}
\frac{d \sigma}{dx dQ^2 dz d p_{h,\perp}^2}  =
\frac{4 \pi \alpha^2}{Q^4 x}
[y^2 2x {\cal F}_1(x,Q^2,z,p_{h,\perp}^2) +
 2(1-y) {\cal F}_2(x,Q^2,z,p_{h,\perp}^2) ]  \; ,
\end{equation}
where $x$, $y$ and $Q^2$ are standard DIS variables, $p_{h,\perp}$ is
the transverse momentum of the detected hadron with respect to the momentum
of the virtual photon and
\begin{equation}
z = \frac{P \cdot p_h}{P \cdot q} \; \stackrel{^{TRF}}{=}
\; \frac{E_h}{\nu}
\end{equation}
is a relativistically invariant variable which in the target rest
frame is the fraction of the virtual photon energy $\nu$ carried by
the hadron.
In the formula above, $P$, $p_h$ and $q$ are the four-momenta of
the target nucleon,
 final hadron and virtual photon, respectively.

\begin{figure}
\mbox{
\epsfysize 3.8cm
\epsfbox{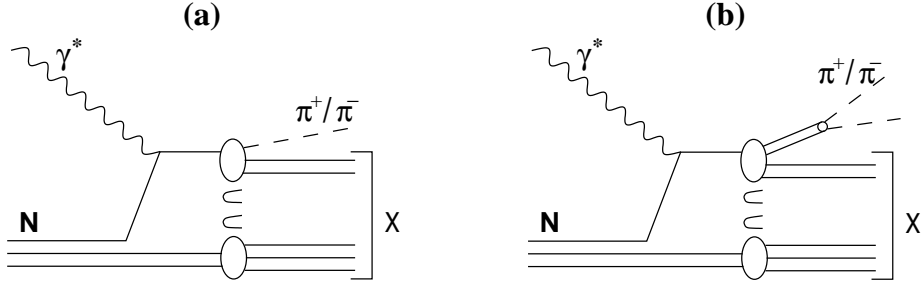}
}
\caption{\it
Partonic mechanisms of pion production:
(a) direct fragmentation of quark into a pion
(b) fragmentation of quark into an intermediate hadronic resonance
and its subsequent decay.}
\label{fig_part_dgs}
\end{figure}

If one is not interested in the transverse momentum distribution of the
emitted hadron then the triple-differential cross section can be
written in a more compact
 way
\begin{equation}
\frac{d \sigma}{dx dQ^2 dz} =
\frac{4 \pi \alpha^2}{Q^4 x}
[y^2 2x {\cal F}_1(x,Q^2,z) +
 2(1-y) {\cal F}_2(x,Q^2,z) ] \; .
\end{equation}
In the quark-parton model (QPM) only mechanisms shown in
Fig.\ref{fig_part_dgs} are assumed.
Usually in calculations
one does not distinguish diagrams (a) and (b).
It is commonly believed that diagram (b) can be included
effectively on the same footing as diagram (a).
We shall discuss later possible restrictions of such an approach.

In the naive QPM the generalized semi-inclusive structure functions
${\cal F}_1$
 and ${\cal F}_2$ are related by
the Callan-Gross relation leaving only one independent structure function,
which can be written as
\begin{equation}
{\cal F}_2^{N \rightarrow \pi}(x,Q^2,z) =
\sum_f e_f^2 x q_f(x,Q^2) \cdot D_{f \rightarrow \pi}(z) \; ,
\label{semi_parton}
\end{equation}
where the sum runs over the quark/antiquark flavours $f=u,d,s$;
$q_f$ are quark distribution functions and $D_{f \rightarrow \pi}(z)$
are so-called fragmentation functions \cite{Feynman_book}.

Quite a 
number of fragmentation functions can be reduced
by the requirement
 of isospin symmetry and charge conjugation:
\begin{eqnarray}
&D_u^{\pi^+}(z) = D_{\bar d}^{\pi^+}(z) =
D_d^{\pi^-}(z) = D_{\bar u}^{\pi^-}(z) & \nonumber \\
& \equiv D_{+}(z) &
\label{favoured_ff}
\end{eqnarray}
for the favoured fragmentation and
\begin{eqnarray}
&D_d^{\pi^+}(z) = D_{\bar u}^{\pi^+}(z) =
D_u^{\pi^-}(z) = D_{\bar d}^{\pi^-}(z) & \nonumber \\
& \equiv D_{-}(z) &
\label{unfavoured_ff}
\end{eqnarray}
for the unfavoured fragmentation.
For the strange/antistrange fragmentation, in principle, a third
type of fragmentation functions has to be assumed. In the following
we shall assume simply:
\begin{equation}
D_s^{\pi^+}(z) = D_{\bar s}^{\pi^+}(z) =
D_s^{\pi^-}(z) = D_{\bar s}^{\pi^-}(z) =
S_{red} \cdot D_{-}(z) \; ,
\end{equation}
where $S_{red}$ is a reduction factor with respect to nonstrange
quarks/antiquarks. In calculations we shall use
$S_{red}$ = 1.0.\footnote{This appears to be not very important in practice.}

Now, in the quark-parton model (\ref{semi_parton}) using symmetry relations 
(\ref{favoured_ff}-\ref{unfavoured_ff}) one can combine semi-inclusive cross 
sections for the production of positive and negative pions on proton and 
neutron targets and isolate a quantity sensitive to the flavour asymmetry 
\cite{HERMES_dbar_ubar}
\begin{equation}
\frac{ \bar d(x) - \bar u(x) }{ u(x) - d(x) } =
\frac{J(z) [1 - r(x,z)] - [1 + r(x,z)]}
     {J(z) [1 - r(x,z)] + [1 + r(x,z)]} \; ,
\label{exp_extract}
\end{equation}
where $J(z) = \frac{3}{5}
\frac{1 + D_{-}(z)/D_{+}(z)}{1 - D_{-}(z)/D_{+}(z)} $
and $r(x,z) = \frac{N_p^{\pi^-}(x,z) - N_n^{\pi^-}(x,z)}
                   {N_p^{\pi^+}(x,z) - N_n^{\pi^+}(x,z)}$
is a ratio of differences of charged pion yields on proton and
neutron.
It is straightforward to see that the fragmentation of strange
quarks/antiquarks cancels in the quantity $r(x,z)$.
It is also worth noting that the r.h.s.\  in Eq.(\ref{exp_extract}),
formally dependent on $z$, gives a quantity independent of $z$.

Thus semi-inclusive production of charged pions in DIS allows us
to determine the asymmetry of light sea quarks. Although this is on
condition that the QPM works well, that is, one may neglect
the influence of other possible mechanisms.

\subsection{Intermediate resonances in the fragmentation}

It is a well known fact that pions produced directly in the fragmentation
process constitute only a fraction of all pions registered in detectors.
The contribution of pions coming from the decay of heavier mesons is of
the same order of magnitude \cite{rho_inclusive_hadron,rho_inclusive_photon,
rho_inclusive_DIS}.

Because the intermediate resonances originate from the fragmentation of the
struck quark their contribution can be included into an overall ``effective''
fragmentation function. Modern analyses of fragmentation functions do
not treat intermediate resonances explicitly, just include them effectively
by fitting total inclusive data.  However, it is not clear, a
priori, whether under a more detailed consideration such an
effective treatment is correct and whether resonances do not disturb
the identity (\ref{exp_extract}).

To treat the intermediate resonances explicitly we
 write down the fragmentation function as a sum of two parts:
a direct fragmentation component (Fig.\ref{fig_part_dgs}a) and
a resonance component (Fig.\ref{fig_part_dgs}b):
\begin{equation}
D_{f \rightarrow \pi} = \tilde D_{f \rightarrow \pi}
                       + \sum_{R} D_{f \rightarrow R \rightarrow \pi} \, ,
\label{ff_decompos}
\end{equation}
where $\tilde D_{f \rightarrow \pi}$ is a fragmentation function of
the direct fragmentation of a quark $f$ into a pion $\pi$,
$ D_{f \rightarrow R \rightarrow \pi} $ describes the production of a pion
$\pi$ through an intermediate resonance $R$ and the sum runs over all
possible resonances. It is known experimentally
that for pion production the vector meson intermediate
states are the most important. For not too small $z$,
neglecting for simplicity transverse momenta, the contribution of the
resonance $R$ to the fragmentation function can be approximated as
\begin{equation}
D_{f \rightarrow R \rightarrow \pi}(z) =
         \int_{z_0}^1 \tilde D_{f \rightarrow R}(z')
            \cdot f_{R \rightarrow \pi} \left( \frac{z}{z'} \right) dz \; ,
\label{res_frag_dec}
\end{equation}
where $z_0 = max(z,z_{min}^{R})$, $z_{min}^{R}$ is the minimal possible
$z$ of the resonance $R$, $\tilde D_{f \rightarrow R}$ is a fragmentation
function for the direct fragmentation of a quark $f$ into a resonance $R$, and
$f_{R \rightarrow \pi}$ describes the decay of the resonance $R$ to
pionic channels.

The fragmentation process transforms quarks with the third component
of isospin \
$
I_3^q = \pm \frac{1}{2} \; \Leftrightarrow \; $ \mbox{\footnotesize
$\left\{ \!
\begin{array}{cc}
u \; \bar d\\
d \; \bar u
\end{array}               \!     \right\} $ } \
into measured pions with
$ \;
I_3^{\pi} = \pm 1 \; \Leftrightarrow \; $ \mbox{\footnotesize
$\left\{ \!
\begin{array}{c}
\pi^+\\
\pi^-
\end{array}               \!     \right\}  \; $ }
i.e.\  there are two initial and two final states of fragmentation with
respect to $I_3$.
If the quark hadronization is driven by the strong interaction
(isospin symmetric) then, in the case of direct fragmentation, one
naturally obtains only two kinds of fragmentation functions related by
(\ref{favoured_ff}-\ref{unfavoured_ff}).
For the resonance contribution
$D_{f \rightarrow \pi}^R \equiv
                       \sum_{R} D_{f \rightarrow R \rightarrow \pi} $,
if the sum comprises \underline{all} possible intermediate states
and if in addition the isospin is conserved in the decay of
resonances, one still has only two kinds of fragmentation functions:

\begin{equation}
D_+^R \equiv \sum_{R} D_{\mbox{
\tiny
$ \left\{ \!
\begin{array}{cc}
u \; \bar d\\
d \; \bar u
\end{array} \!                          \right\}
\rightarrow R \rightarrow  \left\{ \!
\begin{array}{c}
\pi^+\\
\pi^-
\end{array}                      \!     \right\} $ }  }
\end{equation}
and
\begin{equation}
D_-^R \equiv \sum_{R} D_{\mbox{
\tiny
$  \left\{ \!
\begin{array}{cc}
u \; \bar d\\
d \; \bar u
\end{array}     \!                      \right\}
\rightarrow R \rightarrow  \left\{ \!
\begin{array}{c}
\pi^-\\
\pi^+
\end{array}                       \!    \right\} $ }  } \; .
\end{equation}
These functions
correspond uniquely to the standard favoured and unfavoured fragmentation
functions and fulfil the relation
(\ref{favoured_ff}-\ref{unfavoured_ff}) needed to obtain the identity
(\ref{exp_extract}).
Thus we come to the conclusion that intermediate resonances do not violate
Eq.(\ref{exp_extract}) i.e.\ do not disturb the procedure of extraction
of the $\bar d - \bar u$ asymmetry.

However, one cannot avoid completely the explicit treatment
of intermediate resonances when modelling pion spectra.
The fragmentation functions (\ref{ff_decompos}) with the
resonance components (\ref{res_frag_dec}) do not (!) obey the QCD evolution
equation; this is usually ignored in the current literature.
It would be useful to separate out the direct
fragmentation contribution to the fragmentation functions, which has a better
chance of obeying the QCD evolution equations. However, to determine the
fragmentation functions of the direct fragmentation one would
need to perform a combined analysis of fragmentation into pions and into
all other resonances having pionic decay channels. Such an involved
experimental analysis has never yet been done.

\subsection{Choice of fragmentation functions}
\label{choice_ff}

In order to estimate the effect of nonpartonic
components on the extraction of $\bar d - \bar u$ asymmetry
we need to fix fragmentation functions with which the main partonic term 
will be calculated.

Modern parametrizations of fragmentation functions
are fitted mostly to data from $e^+ e^-$ collisions.
These analyses include leading or next-to-leading order QCD corrections
(see for instance \cite{CGGRW94,BKK94,BKK95}).
In contrast to $e^+ e^-$ collisions the situation
in $e p$ scattering is much less developed:
less experimental data, no QCD analysis.

Let us see how the existing parametrizations of fragmentation functions
behave in $e p$ collisions.
We start the quantitative estimations by comparing
the existing fragmentation
 functions with the $z$-distributions
of charged
 pions in $e p$ scattering.
\begin{figure}[h]
\mbox{
\epsfysize 6.0cm
\epsfbox{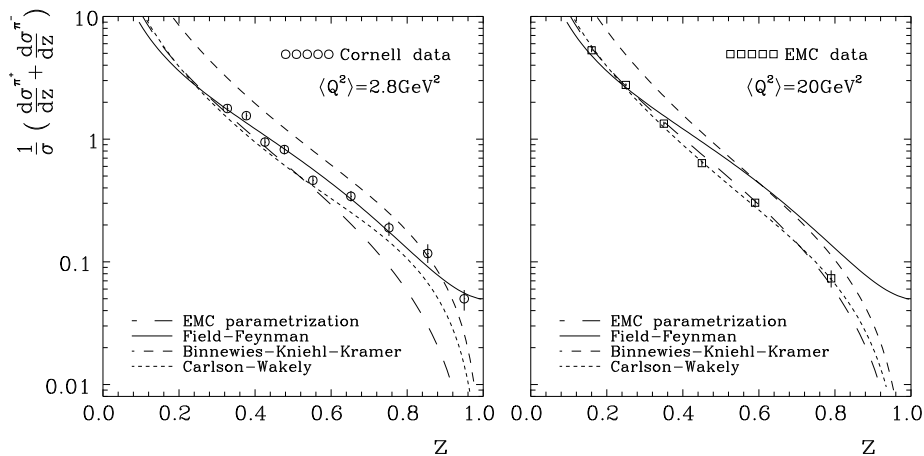}
}
\caption{\it
Multiplicity distribution of charged pions
$\frac{1}{\sigma}
( \frac{d \sigma^{\pi^+}}{dz} + \frac{d \sigma^{\pi^-}}{dz} )$ in DIS.
Different sets of fragmentation functions are confronted with
the Cornell \cite{Drews78} data with 2 $< Q^2 <$ 6 GeV\,$^2$ and
W $\sim$ 3-4 GeV (left panel) and the EMC \cite{EMC85} data
with 20$< \, <\!\! Q^2 \!\!> \, <$71 GeV\,$^2$, depending on $z$ (right panel). }
\label{fig_cha_pi}
\end{figure}
In Fig.\ref{fig_cha_pi} we present
$\frac{1}{\sigma}( \frac{d \sigma^{\pi^+}}{dz} +
                   \frac{d \sigma^{\pi^-}}{dz} )$
data obtained long ago at Cornell \cite{Drews78} (panel a)
with kinematics similar to the HERMES experiment, and the data
from EMC \cite{EMC85} (panel b) with slightly higher $Q^2$.
We show also the QPM predictions obtained with fragmentation functions
from the fit to $e^+ e^-$ data \cite{BKK94},
from the fit to $e^+ e^-$
 and photoproduction data \cite{CW93}
which include QCD corrections, and with fragmentation functions from
the simple QPM fits to the $e p$ data \cite{FF77,EMC85}.
Surprisingly the ``advanced'' parametrizations give a much worse
description of the data than simple ones do. However, simple
parametrizations are limited to the relevant values of $Q^2$.

In principle the modern fragmentation functions used in
$e^+ e^-$ \cite{BKK94} were obtained including QCD corrections,
i.e.\  beyond the naive quark-parton model.
The correct formulae for the cross section in DIS calculated including QCD
corrections are more complicated than QPM ones \cite{BF79,AEMP79}.
On the other hand the analysis of the HERMES experiment
\cite{HERMES_dbar_ubar}
 was performed
based on simple QPM formulae, using Eq.(\ref{exp_extract}).
Therefore in order to compare
 with those results we have to stay
at the QPM level too and ignore some inconsistency. Moreover
the QCD evolution of fragmentation functions (included in the calculation)
changes the pion multiplicity in agreement with the trend of
experimental data and might create the main part of the $Q^2$-dependence
even when used with QPM formulae.
The whole effect of the inconsistency should,
however, be clarified in the future.

In electron-positron scattering the number of negative and
positive pions produced is identical. This is not the case for
$e p$ scattering. Here quark distributions in the proton,
isospin-asymmetric by their nature,
allow us to distinguish between the favoured and unfavoured
fragmentation which is very difficult, if not impossible,
 in $e^+e^-$ scattering.
As can be seen from Eq.(\ref{exp_extract}) such a separation is
essential for our analysis.
\begin{figure}
\begin{center}
\mbox{
\epsfysize 13.0cm
\epsfbox{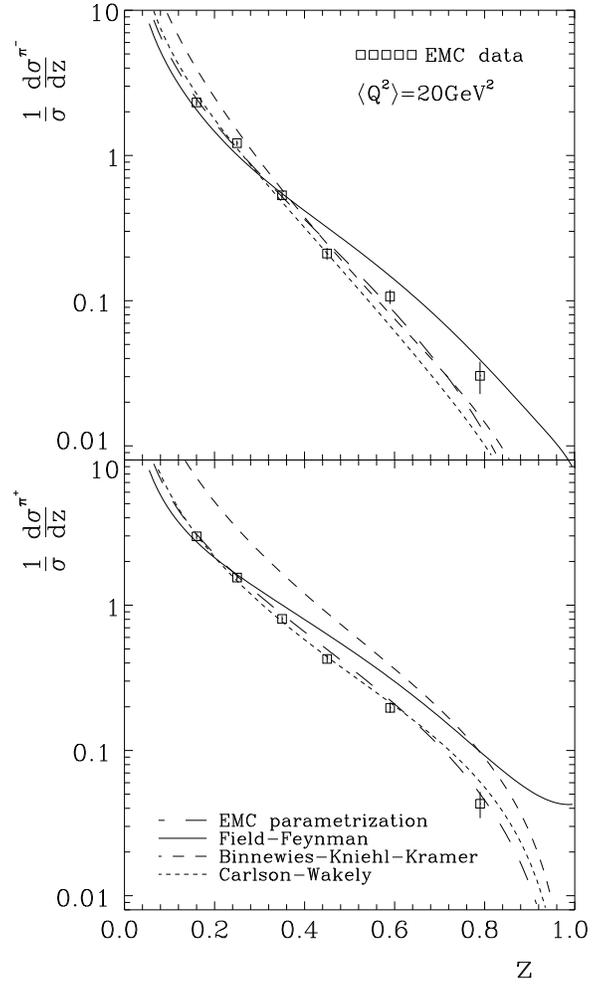}
}
\end{center}
\caption{\it
$\frac{1}{\sigma} ( \frac{d \sigma^{\pi^{-}}}{dz})$ (upper panel)
and
$\frac{1}{\sigma} ( \frac{d \sigma^{\pi^{+}}}{dz})$ (lower panel).
The experimental data are taken from \cite{EMC85}.
The fragmentation functions used are the same as in the previous figure.}
\label{fig_nepo_pi}
\end{figure}
In Fig.\ref{fig_nepo_pi} we show
 $z$-distributions of
negative (upper panel)
 and positive (lower panel)
pions as measured by the EMC
 collaboration \cite{EMC85}.
Different sets of fragmentation functions are confronted with
the experimental data. A surprisingly poor description of the data
is obtained with fragmentation functions from $e^+ e^-$ scattering
\cite{BKK94}. Not very good agreement of the Field-Feynman parametrization
is most probably due to a different $Q^2$ here ($\sim$ 20 GeV$^2$) than
that where it was designed \cite{FF77}.
 A correct QCD evolution
should resolve this disagreement.
As will be discussed below both QPM-parametrizations give
reasonable representations of the ratio of unfavoured
to favoured fragmentation functions, which is a more slowly
QCD-evolving
 quantity.

\begin{figure}[t]
\begin{center}
\mbox{
\epsfysize 7.5cm
\epsfbox{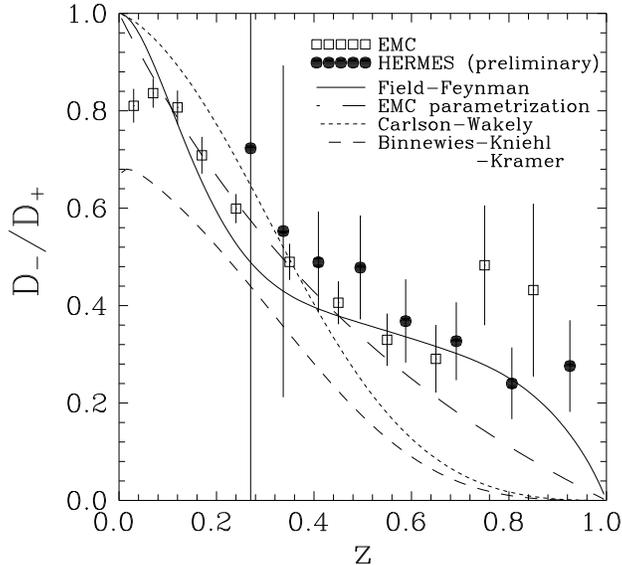}
}
\end{center}
\caption{\it
The ratio of unfavoured to favoured fragmentation functions
as a function of $z$. The experimental data are from
EMC \cite{EMC89} and from a preliminary, unpublished
HERMES analysis \cite{Geiger98}. }
\label{fig_unf_fa_r}
\end{figure}

The ratio $D_{-}(z) / D_{+}(z)$
 directly enters
formula (\ref{exp_extract})
 and, as our analysis shows,
the measured $\bar d - \bar u$
asymmetry is very sensitive to it.
In Fig.\ref{fig_unf_fa_r} we display the ratio
obtained from the analysis of experimental data from
EMC \cite{EMC89} and recently obtained
 by the HERMES
collaboration at DESY \cite{Geiger98}.
The simple QPM
 parametrizations \cite{FF77,EMC85} provide a reasonable
description of the data.
In contrast the``advanced''
fragmentation functions
\cite{BKK94,CW93} fail again.
Is it due to a different physics in $e^+e^-$ collisions than in DIS,
or is it due to QCD corrections, or is it due to something else?
In our opinion this
is mainly due to the fact that
the $e^+e^-$ data are not sufficient to separate unambiguously
the favoured and unfavoured fragmentation functions.
A QCD analysis of DIS fragmentation functions is called for.

The analysis above advocates the Field-Feynman
parametrization \cite{FF77} as the only good representation of
the available $e p$ data in the HERMES kinematical region.
This parametrization will be used in the following analysis.

\section{Nonpartonic components}

\begin{figure}[t]
\begin{center}
\mbox{
\epsfysize 9.0cm
\epsfbox{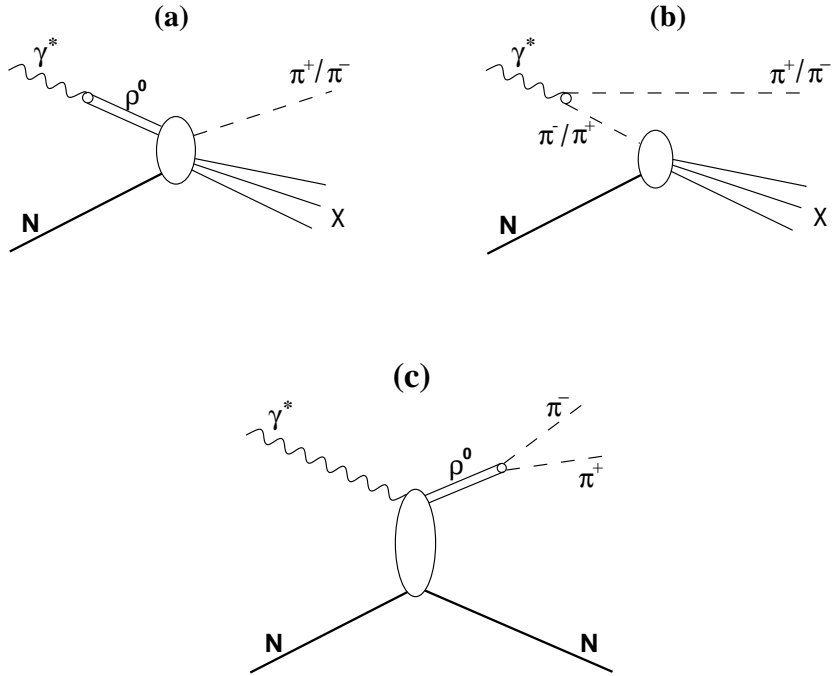}
}
\end{center}
\caption{\it
Nonpartonic mechanisms of pion production taken into account in this work:
(a) VDM contribution,
(b) spectator mechanism,
(c) elastic production of the $\rho^0$ meson and its decay.}
\label{fig_nonp_dgs}
\end{figure}

For small $Q^2$, as for the HERMES experiment, some mechanisms
of nonpartonic origin (see e.g.\  Fig.\ref{fig_nonp_dgs})
may become important.
For instance, the virtual photon can interact with
the nucleon via its intermediate hadronic state. Such a mechanism is
usually described within the vector dominance model (VDM).
The photon could also fluctuate into a pair of pions, where
both or one of them interact with the nucleon. In addition
some exclusive processes can produce pions directly or as decay
products of heavier mesons.

To our best knowledge none of such processes has been investigated
in the literature. Their influence on the extracted $\bar d - \bar u$
asymmetry also remains unknown. We shall discuss processes shown
in Fig.\ref{fig_nonp_dgs}
 one by one.

\subsection{Central VDM contribution}

Let us start from the VDM component (see Fig.\ref{fig_nonp_dgs}a).
It was shown that in the inclusive DIS incorporation of the VDM contribution
and related modification of the partonic component help
to understand the behaviour of structure functions $F_2^p$ and $F_2^d$
at small $Q^2$ \cite{BK_model,SU_inclusive}.
This model was confirmed by a recent
analysis of the $Q^2$ dependence of the world data for the structure
function difference $F_2^p - F_2^n$ \cite{SU_part_viol}.

The model for inclusive structure functions
\cite{BK_model,SU_inclusive} can be generalized
to semi-inclusive production of pions:
\begin{eqnarray}
{\cal F}_2^{N \rightarrow \pi}(x,Q^2,z) &=&
\frac{Q^2}{Q^2+Q_0^2}
\sum_f e_f^2 x q_f(x,Q^2) \cdot D_{f \rightarrow \pi}(z) \nonumber \\
&+& \frac{Q^2}{\pi}
\sum_{V} \frac{1}{\gamma_V^2}
\frac{\sigma_{VN \rightarrow \pi X}(s^{1/2}) M_V^4}{(Q^2+M_V^2)^2}
\Omega_V(x,Q^2)
\; .
\label{SU_semi}
\end{eqnarray}
The second sum above runs over vector mesons $V = \rho^0, \omega, \phi$
and
 $\Omega_V$ is a correction factor which takes into account
finite fluctuation times
 of the virtual photon into vector mesons for
large $x$ \cite{SU_inclusive}.

The inclusive cross section for pion production in vector meson
($\rho^0, \omega, \phi$) scattering off the proton and neutron is not
known experimentally. There is no model in the literature that one can trust
quantitatively but
in analogy to the total $\rho^0 N$ cross section the
cross section
$\rho^0 N \rightarrow \pi^{\pm} X$ can be estimated as:
\begin{eqnarray}
\sigma(\rho^0 p \rightarrow \pi^{\pm} X) &\approx&
 1/2 \; [ \sigma(\pi^+ p \rightarrow \pi^{\pm} X)
         + \sigma(\pi^- p \rightarrow \pi^{\pm} X)  ]
\nonumber \\
\sigma(\rho^0 n \rightarrow \pi^{\pm} X) &\approx&
  1/2 \; [ \sigma(\pi^+ n \rightarrow \pi^{\pm} X)
         + \sigma(\pi^- n \rightarrow \pi^{\pm} X) ]
\; .
\end{eqnarray}

Experimental data from the ABBCCHW collaboration \cite{Bosetti73} at
$p_{lab}^{\pi}$ = 8, 16 GeV correspond approximately to the range
of the HERMES experiment \cite{HERMES_dbar_ubar}.
Unfortunately as it often happens in high-energy physics there is only data
for proton targets. Using isospin symmetry for hadronic reactions one can
obtain corresponding cross sections on the neutron from those on the proton
by assuming
\begin{eqnarray}
\sigma(\rho^0 n \rightarrow \pi^+X) &=&
\sigma(\rho^0 p \rightarrow \pi^-X) \; , \\
\sigma(\rho^0 n \rightarrow \pi^-X) &=&
\sigma(\rho^0 p \rightarrow \pi^+X) \; .
\label{vdm_isosp_sym}
\end{eqnarray}
These relations hold not only for the total cross sections
but also for differential ones independently of energy.
From the most complete data at $p_{lab}^{\pi}$ = 16 GeV \cite{Bosetti73}
(W = 5.56 GeV) we get \\
$ 1/2 \; \left[ \, \sigma(\pi^+ p \rightarrow \pi^+ X) +
      \sigma(\pi^- p \rightarrow \pi^+ X) \, \right] $
             = 38.65 $\pm$ 0.29 mb ,\\
$ 1/2 \; [\, \sigma(\pi^+ p \rightarrow \pi^- X) +
         \sigma(\pi^- p \rightarrow \pi^- X)$ ] = 31.80 $\pm$ 0.22 mb ;\\
clearly different values.
Although the bulk of the difference comes from the target fragmentation
region, which we are not interested in, in the beam fragmentation
region it was also found $\sigma(\pi^+p \rightarrow \pi^- X)$ = 14.8 mb and
$\sigma(\pi^-p \rightarrow \pi^+ X)$ = 19.0 mb \cite{Bosetti73} with
almost equal cross sections for beam-like pions.

The situation in the semi-inclusive case is more complicated
than for total cross sections.
The experimental spectra for $\pi^{\pm} p \rightarrow \pi^{\pm} X$
contain components due to peripheral processes, which are,
in general, specific, different for different reactions.
Peripheral processes from the
$\pi^+ p \rightarrow \pi^+ X$ and $\pi^- p \rightarrow \pi^- X$
reactions do not contribute to the
$\rho^0 p \rightarrow \pi^{\pm} X$ reaction and should be eliminated;
only nondiffractive components of the $\pi p \rightarrow \pi X$
 reactions
 should be taken into account when modelling $\rho^0$-induced
 reactions.\footnote{
Some peripheral processes specific for the $\rho^0$ beam
will be included explicitly in section \ref{sec_excl_rho}.}
This requires physically motivated parametrization
of the $ \pi N \rightarrow \pi X$ data.

Following these arguments we have parametrized the experimental differential
cross sections for four different reactions
$\pi^{\pm} p \rightarrow \pi^{\pm} X$ from \cite{Bosetti73} as
a sum of central and peripheral components
\begin{equation}
\frac{d \sigma}{dx_F dp_{\perp}^2} =
\frac{d \sigma^{cen}}{dx_F dp_{\perp}^2} +
\frac{d \sigma^{per}}{dx_F dp_{\perp}^2}
\; ,
\end{equation}
where $x_F$ is the well known Feynman variable.
Details of this analysis will be presented elsewhere,
a short sketch is given in Appendix \ref{app_bosetti}.
Because the CM-energy of the ABBCCHW collaboration is very similar
to that of the HERMES experiment,
we believe that in the range of energy relevant for this experiment
the functional form given in the appendix is suitable.

Finally, our analysis of experimental data \cite{Bosetti73}
combined with the assumption of isospin symmetry (\ref{vdm_isosp_sym})
shows that for the nondiffractive components still
\begin{equation}
\sigma(\rho^0 p \rightarrow \pi^{\pm} X) \ne
\sigma(\rho^0 n \rightarrow \pi^{\pm} X) \; .
\end{equation}
%
\begin{figure}[t]
\begin{center}
\mbox{
\epsfysize 8.0cm
\epsfbox{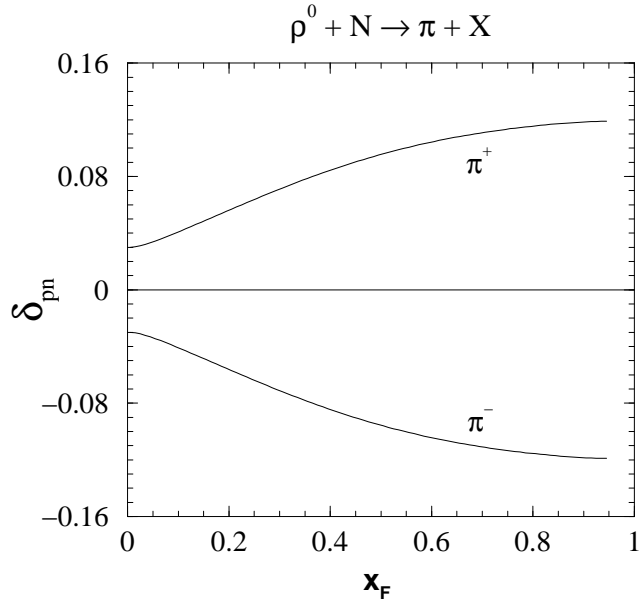}
}
\end{center}
\caption{\it
$\delta_{pn}^{\rho^0 \rightarrow \pi^{\pm}}$ calculated from
Eq.(\ref{VDM_pn_asymmetry}) as a function of Feynman $x_F$.
The pion+proton cross sections were taken
from the parametrization described in Appendix \ref{app_bosetti}.}
\label{fig_apn_vdm}
\end{figure}
In Fig.\ref{fig_apn_vdm} we show
\begin{equation}
\delta_{pn}^{\rho^0 \rightarrow \pi^{\pm} } = \frac
{\sigma(\rho^0 p \rightarrow \pi^{\pm} X)
          - \sigma(\rho^0 n \rightarrow \pi^{\pm} X) }
{\sigma(\rho^0 p \rightarrow \pi^{\pm} X)
          + \sigma(\rho^0 n \rightarrow \pi^{\pm} X) }
\label{VDM_pn_asymmetry}
\end{equation}
for both positive and negative pions. An identical value of the asymmetry
is expected
 for the $\omega$ beam and a similar one for the $\phi$ beam.
This automatically means practically the same result for photon (real or
virtual) induced
reactions which proceed via hadronic intermediate states.  We obtain rather
large asymmetries, larger than for total photoproduction
cross sections on the proton and neutron.

Different cross sections on proton and neutron mean that the VDM
contribution modifies the r.h.s.\ of Eq.(\ref{exp_extract}).
\begin{figure}[tp]
\begin{center}
\mbox{
\epsfysize 9.0cm
\epsfbox{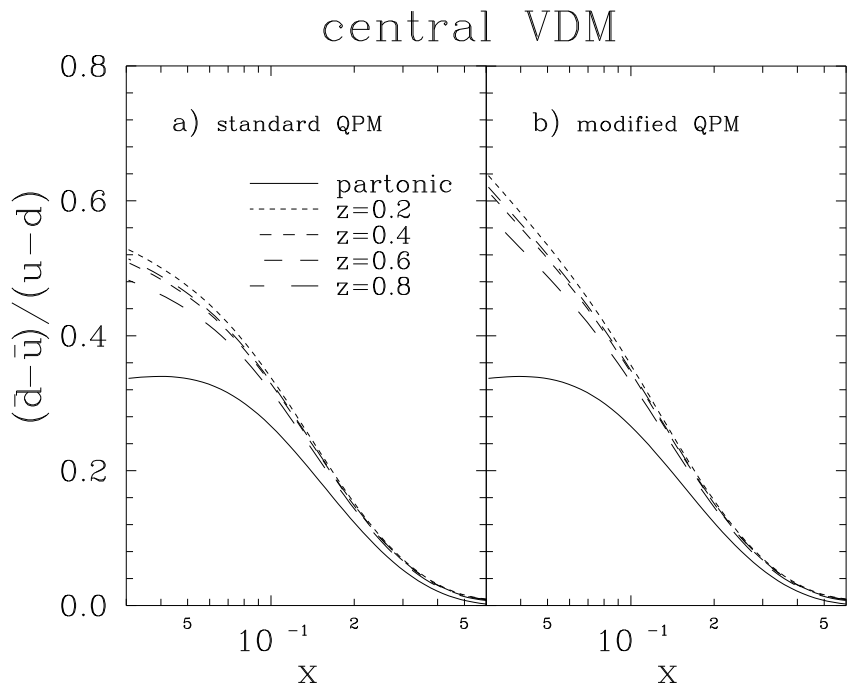}
}
\end{center}
\caption{\it
The true (solid) and the modified by the central VDM contribution
$\frac{\bar d - \bar u}{u - d}$, calculated according to the l.h.s. and r.h.s.
of Eq.(\ref{exp_extract}), respectively, as a function of Bjorken-$x$
for different values of $z$ and typical HERMES $W$ = 5 GeV.
The result in panel (a) is obtained by a simple addition of the fragmentation
and the central VDM components ($Q^2_0$ = 0 in Eq.(\ref{SU_semi})).
The result in panel (b) is obtained with the rescaled
($Q^2_0$ = 0.8 GeV\,$^2$)
fragmentation component, as described in the text.}
\label{fig_hermes_v}
\end{figure}
In Fig.\ref{fig_hermes_v} we show a modification of the measured quantity
$\frac{\bar d - \bar u}{u - d}$ due to the central VDM component.
In the HERMES experiment \cite{HERMES_dbar_ubar} both $Q^2$ and $W$
vary with Bjorken-$x$, but the change of energy is considerably smaller.
In the present calculation the photon-proton CM energy was fixed at
the average HERMES value $W$ = 5.0 GeV$^2$.
In the panel (a) we show a result of a calculation where the central
 VDM component
discussed in this section is simply added to the main fragmentation component.
 In the panel (b) the fragmentation component in addition
was rescaled by the factor
 $\frac{Q^2}{Q^2+Q_0^2}$ (see Eq.(\ref{SU_semi}))
which is required for consistency of inclusive and
semi-inclusive structure
 functions.
The solid line in both panels represents $\frac{\bar d - \bar u}{u - d}$
obtained directly from the parton distributions \cite{GRV94}.
As can be seen from the figure the r.h.s.\ of Eq.(\ref{exp_extract}) clearly
deviates from the assumed partonic outcome.
The effect is surprisingly large,
 especially
 for small $x$.\footnote{
We wish to remind the reader here that in the HERMES experiment
 the photon virtuality
for small $x$ is only of the order of 1 - 2 GeV$^2$.}
Thus, the quark flavour asymmetry
 extracted from semi-inclusive
experiments in the simple QPM approach seems to be highly overestimated
if the VDM contribution is neglected.

The VDM effect discussed in this section is not completely new.
A similar effect of the hadronic structure of the photon
on the difference of semi-inclusive cross sections
$\sigma_{\gamma p \rightarrow \pi^+ X} -
 \sigma_{\gamma p \rightarrow \pi^- X}$
was already noticed long ago in real photoproduction \cite{FMSP80}.
Although in DIS the effect is smaller, it, however, strongly modifies
the measured $\bar d - \bar u$ asymmetry.

\subsection{Spectator mechanism}

In both partonic and central VDM mechanisms the virtual photon is
totally absorbed and pions are produced in a complex process
involving many degrees of freedom. Such pions are then preferentially
emitted at not very large values of $z$. The peripheral processes
are not included either in the partonic or the central VDM
component considered above and are expected to be important in the region of large
$z$ where these processes disappear.
Let us consider first the spectator mechanism depicted
in Fig.\ref{fig_nonp_dgs}b. To our knowledge such a mechanism
has been never discussed in the literature for photon induced
reactions. We begin with the case of real photoproduction
where we can apply a technique from \cite{SHS96}.
For virtual photons  with $q^2 < 0$ the formalism is not well developed.

The cross section for the spectator pions
$\pi^{\pm}$ in real photoproduction can be expressed
as a product of
the distribution of pions in the photon
($f_{\pi^{\pm}/\gamma}$) and the off-shell total cross section for scattering of
$\pi^{\pm}$ off the proton or neutron:
\begin{equation}
\frac{d \sigma_{spect}^{\pi^{\pm}}}{dz} \approx
f_{\pi^{\pm}/\gamma} (z) \cdot
\sigma_{tot}^{\pi^{\mp} N} ((1-z)s) \; .
\label{spectator}
\end{equation}
At small $\pi N$ energies
$s_{\pi N} \approx (1-z) s_{\gamma^* N}$ relevant for the spectator
mechanism there can be a difference between $\sigma_{tot}^{\pi^+p}$
and $\sigma_{tot}^{\pi^-p}$. Together with isospin symmetry of hadronic
reactions this would lead to different
$N^{\pi^\pm}_p$ and $N^{\pi^\pm}_n$ i.e.\ would modify the identity
(\ref{exp_extract}). To represent the total cross sections of
$\pi^{\pm}$ scattering off nucleons we use cubic interpolation of
the world experimental data \cite{PDB98}.

The distribution function $f_{\pi^{\pm}/\gamma}$
can be calculated using a technique similar to that outlined in
\cite{SHS96}:
\begin{equation}
f_{\pi^{\pm}/\gamma}(z,k_{\perp}^2) = \frac{g_{em}^2}{64 \pi^2}
\cdot
\frac{1}{z(1-z)} \cdot
\frac{2 k_{\perp}^2}{[0 - M_{\pi \pi}^2(z,k_{\perp}^2)]^2}
|F(z,k_{\perp}^2)|^2  \; ,
\label{lc_flux}
\end{equation}
where $g_{em}$ is the electromagnetic $\gamma \rightarrow \pi^+ \pi^-$
coupling constant, $M_{\pi \pi}$ is the invariant mass of
the two-pion system, $F(z,k_{\perp}^2)$ is a vertex form factor which
accounts for the finite size of particles involved and off-shell effects.
For other technical details see \cite{SHS96}.

\begin{figure} 
\begin{center}
\mbox{
\epsfysize 7.5cm
\epsfbox{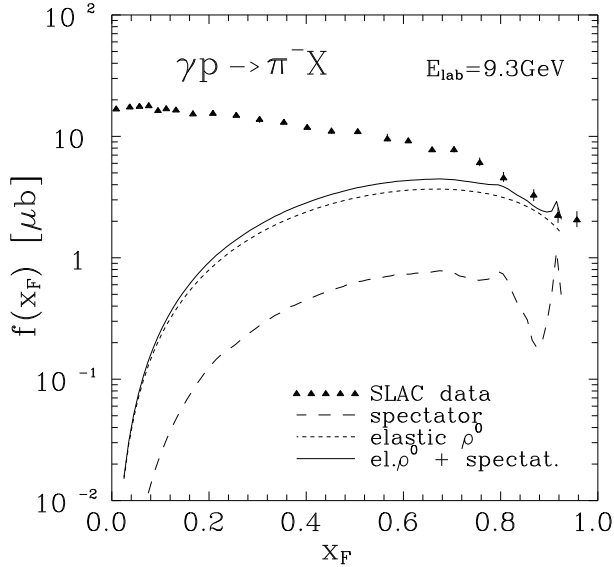}
}
\end{center}
\caption{\it
The invariant structure function $f(x_F)$ for the
$\gamma p \rightarrow \pi^-X$ reaction and $E_{\gamma}$ = 9.3 GeV.
}
\label{fig_photo_pi}
\end{figure}
In Fig.\ref{fig_photo_pi} we show the result of our calculation
(dashed line) with cut-off mass 1.5 GeV as in Ref.\cite{SHS96} for
$s^{1/2}$ = 4.28 GeV corresponding to the experimental data from SLAC
\cite{Moffeit72} with $E_{\gamma}$ = 9.3 GeV. The results are presented
in terms of the invariant single particle structure function \cite{Moffeit72}
(see also Eq.(\ref{inv_spstrf}) in Appendix \ref{app_bosetti}).
The energy in the pion-proton subsystem decreases with increasing $x_F$
and we observe fluctuations due to $s$-channel $\pi N$ resonances.
The peaks would be even more pronounced at smaller photon energies
and would disappear completely at larger photon energies.

The contribution of the spectator mechanism calculated with reasonable
cut-off masses in the vertex form factor is smaller than
experimental data.
A dominant fraction of pions produced at large $x_F$ appears to be
given by the mechanism of elastic $\rho^0$-meson production and its
subsequent decay (discussed in detail in section \ref{sec_excl_rho}).
By the short-dashed line we present this contribution corresponding to
the cross section calculated as:
\begin{equation}
\frac{d \sigma}{dz}^{ \rho^0 \rightarrow \pi^{\pm} }
= \sigma(\gamma p \rightarrow \rho^0 p)
\cdot
f_{decay}^{\rho^0 \rightarrow \pi^{\pm}}(z) \; ,
\end{equation}
where the explicit functional form of the decay function
$f_{decay}^{\rho^0 \rightarrow \pi^{\pm}}(z)$ can be found in
Appendix \ref{app_rho0_dec}.
In the calculation we have used the experimental value
$\sigma(\gamma p \rightarrow \rho^0 p)$ = 13 $\mu$b, relevant
for $E_{\gamma}$ = 9.3 GeV. The sum of both contributions
(solid line) seems to be consistent with the large-$x_F$ part of
the pion spectrum.
Thus in real photoproduction the spectator mechanism becomes
non-negligible only at large $x_F$ (or $z$).

In virtual photoproduction the situation is more complicated.
The absolute normalization of $f_{\pi^{\pm}/\gamma^*}$ should
depend on the virtuality of the photon.
However, it cannot be calculated from first principles as it involves
a vertex form factor where more than one particle is off mass shell.
Empirically such cases are strongly damped \cite{NSCHS98}.
Also a naive use of a typical light-cone parametrization of the
vertex
form factors \cite{SHS96} with the mass of the parent particle replaced
by the virtuality of the photon leads to a strong suppression in
comparison to real photoproduction. How strong this suppression is
in comparison to the suppression of the elastic-$\rho^0$ production
mechanism is not clear. For photon virtualities of the order
of $Q^2 \sim$ 2-4 GeV$^2$ we find that the elastic-$\rho^0$
contribution, discussed in detail in the next section, together with
the partonic component, totally accounts for the cross sections at large $z$,
leaving practically no room for the spectator mechanism
(see Fig.\ref{fig_mult_all}).

In summary , the spectator mechanism, while potentially
important in real photoproduction is most probably negligible in
DIS.

\subsection{Exclusive $\rho$ meson production}
\label{sec_excl_rho}

The exclusive meson production
$\gamma^* N \rightarrow M N'$ is one more mechanism not included
in the fragmentation formalism (Eq.(\ref{semi_parton})) and
may also modify the extraction of the $\bar d - \bar u$ asymmetry.
The pion exclusive channels ($M = \pi$)\footnote{
These are not included in the spectator mechanism discussed
above where the final state $X = N$ was not taken into account.}
 contribute at $ z \approx$ 1,
i.e.\ outside of the range of the HERMES kinematics
and will be ignored in the following discussion.
In contrast, the pions from the decays of light vector mesons may be
important in the context of the $\bar d - \bar u$ asymmetry from
semi-inclusive pion production.
The production of $\rho$ mesons ($M = \rho$) seems to be of
particular importance. Firstly, the $\rho^0 N$ channel is known to
be the dominant exclusive channel in $\gamma^* N$ scattering.
Secondly, because $\rho^0$ decays predominantly
into two pions this will
produce pions with $< \! z \! > \, \sim \frac{1}{2}$. A detailed calculation
(see Appendix \ref{app_rho0_dec}) shows that the dispersion of the decay pion
$z$-distribution is large and therefore this effect has a chance
of being observed at large $z$ where the hadronization rate is already
much smaller. The next potentially important mechanism is the $\omega$
meson ($M = \omega$) production. However, the dominant $\omega$ meson
decay channel is the  three-body system $\pi^+ \pi^- \pi^0$ i.e.
it is expected to contribute to the inclusive pion distribution at
$< \! z \! > \, \sim \frac{1}{3}$, i.e.\  in the region where the
hadronization
rate is large. Moreover the cross section for the $\omega$ channel
is smaller than for the $\rho^0$ channel.
Below we shall consider
the $\rho^0$ channel only, which is probably the most important.

The elastic $\rho^0$-production contribution
(diagram (c) in Fig.\ref{fig_nonp_dgs}) to the
semi-inclusive structure function (\ref{semi_parton},\ref{SU_semi})
can be written formally as
\begin{equation}
{\cal F}_2^{\rho^0,el}(x,Q^2,z) = \frac{Q^2}{4 \pi^2 \alpha}
\cdot \sigma_{\gamma^* N \rightarrow \rho^0 N}(W,Q^2) \cdot
f_{\rho^0 \rightarrow \pi}(z)  \; ,
\end{equation}
The decay function $f_{\rho^0 \rightarrow \pi}$ can be easily
calculated (see Appendix \ref{app_rho0_dec}).
For not too high energies, as for the HERMES experiment
one may expect
$\sigma(\gamma^* p \rightarrow \rho^0 p) \ne
 \sigma(\gamma^* n \rightarrow \rho^0 n)$ which would modify
the $\bar d - \bar u$ asymmetry.
At high energy the pomeron-exchange (two-gluon exchange)
mechanism dominates and one may expect
$\sigma(\gamma^* p \rightarrow \rho^0 p) =
 \sigma(\gamma^* n \rightarrow \rho^0 n)$.
At low energy the exchange of subleading reggeons (quark
exchange) could lead to
$\sigma(\gamma^* p \rightarrow \rho^0 p) \ne
 \sigma(\gamma^* n \rightarrow \rho^0 n)$ due to isovector
contributions. In real photoproduction the isovector amplitude is known
to be rather small \cite{Leith78}. In DIS the situation may
be quite different.
Assuming that the production is hard, i.e.\ of perturbative nature,
the longitudinal $\rho^0$ is predicted to be dominated by
the quark exchange mechanism at low photon energies \cite{VGG98}.
The HERMES $\gamma^* p$ energy
corresponds precisely to the maximum of the $\rho_L^0$ production in
the hard quark-exchange exclusive reaction $\gamma^* p \rightarrow
\rho_L^0 p$ \cite{VGG98,VGG99,MPW98,MPW99}.
Although there are no experimental data in this region,
the data from EMC \cite{EMC_rho0}, NMC \cite{NMC_rho0} and
E665 \cite{E665_rho0} collaborations in the close neighbourhood
seem to be in rough agreement with the calculation for $\sigma_L$
\cite{VGG98,VGG99}.
Different quark distributions in the proton and neutron lead
in this approach to different $\rho^0$ production cross sections
for proton and neutron targets, which
obviously leads to a different production
rate of charged pions in reactions on the proton and neutron.
This in turn modifies the right hand side of Eq.(\ref{exp_extract}) and
the subsequent conclusions on the $\bar d - \bar u$ asymmetry.

One could also try to understand elastic meson production
within the Regge phenomenology \cite{DL_rho0,HKK_rho0}.
It is not obvious a priori
what is the kinematical range of applicability
of either the quark exchange approach or Regge phenomenology.
Below we shall investigate elastic $\rho^0$ production on the proton and
neutron using both these approaches:\\
\quad a) the Regge approach, and\\
\quad b) a QCD inspired quark-exchange model.

\subsubsection{Regge approach}

\begin{figure}[t]
\begin{center}
\mbox{
\epsfysize 9.3cm
\epsfbox{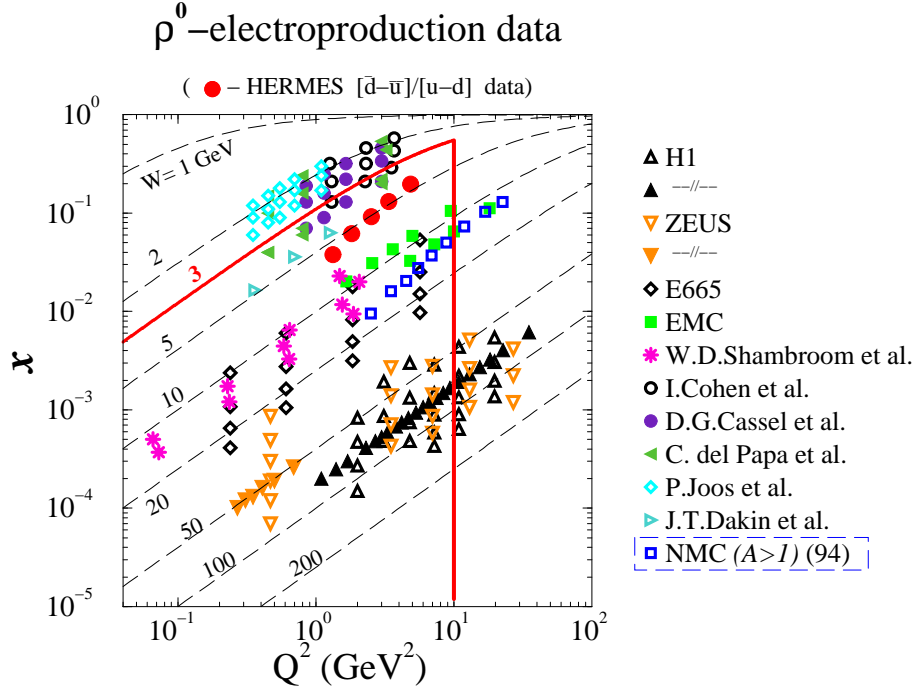}
}
\end{center}
\caption{\it
Experimental data points for exclusive $\rho^0$ production on the
proton and deuteron (open squares). For illustration the HERMES
kinematics is shown in addition by large solid circles.
The thick solid lines show the limits for the Regge-inspired fit.}
\label{fig_map_txt}
\end{figure}

The cross section for neutral $\rho$ meson electroproduction in the
HERMES kinematics is not known experimentally. While for the proton
target there are data in slightly different kinematical regions of $x$
and $Q^2$ \cite{proton_rho0}, there is almost no data for the
neutron target.
Only in one work \cite{E665_pd_rho0} was the $\rho^0$ production
studied simultaneously for the proton and deuteron data.
The $x$- and $Q^2$-integrated result obtained there\\
\begin{center}
$\sigma_{incoher}(\gamma^* d \rightarrow \rho^0 pn) =
((0.7 - 0.8) \pm 0.2) \cdot
\sigma(\gamma^* p \rightarrow \rho^0 p)$ \\
\end{center}
does not exclude the difference between the proton and neutron target.
If the nuclear effects are completely ignored this leads to \\
\begin{center}
$\sigma(\gamma^* n \rightarrow \rho^0 n) =
((0.4 - 0.6) \pm 0.4) \cdot
\sigma(\gamma^* p \rightarrow \rho^0 p)$.\\
\end{center}
However, in the present analysis we are interested in $x$- and
$Q^2$-dependent cross sections.
The only electroproduction data on the deuteron target with well
defined kinematics were published in \cite{NMC_rho0}.
Such a limited set of the deuteron data does not allow us
to determine the cross section on the neutron target for $x$ and
$Q^2$ in the kinematical region we need.
A possible way out would be to parametrize the proton data
with a suitable, physically motivated parametrization
in a possibly broad kinematical range (there are rich data around the
kinematical region of the HERMES experiment, see Fig.\ref{fig_map_txt}) and
use theoretical arguments and/or experimental data for other reactions
to determine the neutron cross sections.

In the present paper we shall parametrize the existing experimental
data for exclusive $\rho^0$ production by means of the following simple
 Regge-inspired reaction amplitude, similar to that in Ref.\cite{DL_rho0},
\begin{eqnarray}
A_{ \lambda_N' \leftarrow \lambda_{N} }
 ^{ \lambda_{V} \leftarrow \lambda_{\gamma^*} }
(\gamma^* N \rightarrow \rho^0 N; t) &=&
\left\{
i \cdot C_{\Pom}(t)
\left( \frac{s}{s_0} \right)^{ \mbox{\large$\epsilon$}_{\mbox{\tiny $\Pom$}} } +
\left[ \frac{-1+i}{\sqrt{2}} \right]
\right.
\cdot C_{IS}(t) \left( \frac{s}{s_0} \right)^{-1/2}
\nonumber \\
&\pm&
\left.
\left[ \frac{-1+i}{\sqrt{2}} \right]
\cdot C_{IV}(t) \left( \frac{s}{s_0} \right)^{-1/2}
\right\}
\nonumber \\
&&\cdot \left[ \frac{m_{\rho}^2}{m_{\rho}^2+Q^2} \right]
\delta_{\lambda_{N'} \lambda_{N}} \delta_{\lambda_{V}
\lambda_{\gamma^*}}
\label{Regge_amplitude}
\end{eqnarray}
with $``+"$ and $``-"$ in front of the isovector ($IV$) contribution
for proton and neutron, respectively. The pomeron contribution is
marked by $\Pom$ and isoscalar reggeon contribution by $IS$.
The following normalization is assumed:
\begin{equation}
\frac{d \sigma}{dt} = \frac{1}{(2 s_N + 1) N_{\lambda}}
 \sum_{\substack{ \lambda_N \lambda_N'\\  \lambda_{\gamma^*} \lambda_V
}}
 \left| A_{\lambda_N' \leftarrow \lambda_N}
         ^{\lambda_V \leftarrow \lambda_{\gamma^*} }(t) \right|^2  \; ,
\label{amplitude_normalization}
\end{equation}
where $s_N$ is the spin of the nucleon and $N_{\lambda}$ is the number
of active helicity states of the virtual photon.

In the following we are interested in relatively low $\gamma^* N$-energies
where in principle the pion-exchange mechanism could be important too.
At low energies, just above resonances, the pion-exchange
mechanism is known to be the dominant mechanism for photoproduction of
 $\omega$ mesons
\cite{Joos77,TO98}. It can be shown that due to the helicity structure of
its amplitude the pion-exchange contribution does not interfere with
the diffractive contribution as far as the spin-averaged cross section
is considered, that is the pion-exchange contribution can be added
incoherently in the cross section.
 Because $\Gamma_{\rho^0 \rightarrow \pi^0 + \gamma}
\ll \Gamma_{\omega \rightarrow \pi^0 + \gamma}$ the relevant coupling
constant
$f_{\gamma \rho^0 \pi}^2 = 96 \pi
 \left(
\frac{m_{\rho^2}}{m_{\rho}^2 - m_{\pi}^2}
 \right)
\Gamma_{\rho \pi \gamma}$
is rather small.
In order to estimate the corresponding cross section one has to make
some plausible estimation for the vertex form factors.
Assuming the form factors which lead to a good description of
the $\omega$-photoproduction data, whilst neglecting other mechanisms,
provides a reasonable upper estimate of the pion-exchange contribution
for $\rho^0$ production.
The pion-exchange contribution estimated in this way
can most probably be neglected in the kinematical region considered
here. It is interesting to notice that unlike for the neutral $\rho$
meson production the pion-exchange contribution cannot
be neglected for charged $\rho$ meson production due to lack of the
dominant isoscalar contributions.
The discussion above further justifies the simple Ansatz used in
(\ref{Regge_amplitude}).

In practical application we assume the same $t$-dependence of
$C_{\Pom}$, $C_{IS}$ and $C_{IV}$ and take $\Lambda = m_{\rho}$.
The total $\gamma^* N \rightarrow \rho^0 N$ cross section can be
obtained as the integral
\begin{equation}
\sigma(W,Q^2) = \int_{t_{min}(W,Q^2)}^{t_{max}(W,Q^2)}
 \; \frac{d \sigma}{dt}(t)\; dt \; .
\label{total}
\end{equation}
There are many approximate or even incorrect formulas
for the upper and lower integration limits in the literature.
It is, however, essential to use correct formulas for small $W$.

\begin{figure}
\begin{center}
\mbox{
\epsfysize 15.1cm
\epsfbox{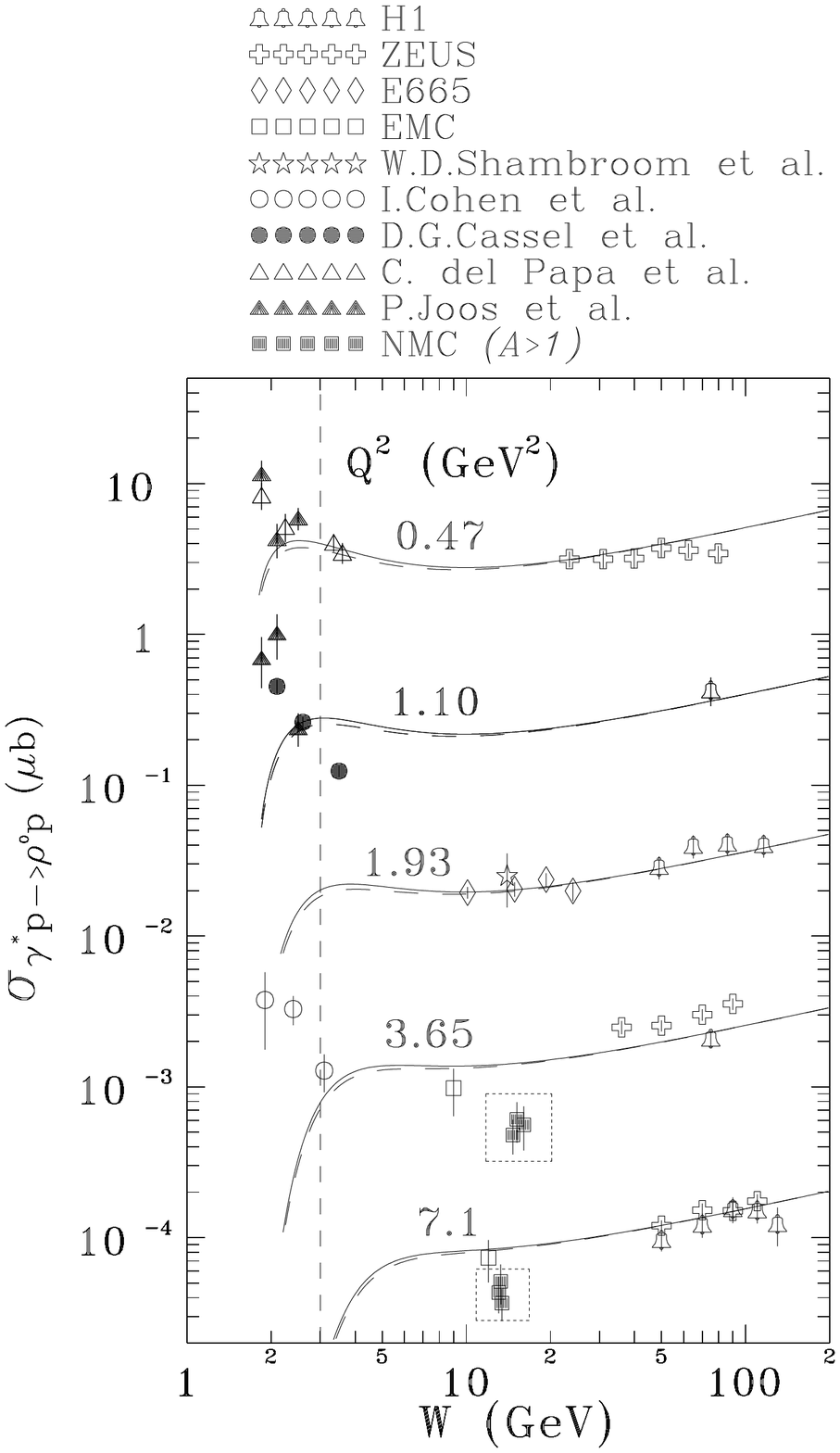}
}
\end{center}
\caption{\it
The cross section for $\gamma^* p \rightarrow \rho^0 p$
as a function of center-of-mass energy for selected values
of photon virtuality. The solid line is obtained from the VDM-Regge
inspired fit. The dased line shows the cross section on the neutron target.
The solid squares represent the NMC nuclear data \cite{NMC_rho0}.
Please note that excepting $Q^2$ = 0.47, all other curves and
experimental points are rescaled by 5, 5$^2$, 5$^3$ and 5$^4$.}
\label{fig_sg_r0ex}
\end{figure}

The free parameters in Eq.(\ref{Regge_amplitude})
i.e.\  $\mbox{\large$\epsilon$}_{\mbox{\tiny $\Pom$}}$,
$C_{\Pom}$ and $C_{IS}+C_{IV}$
\footnote{It is impossible to separate the isoscalar and isovector
reggeon contributions from the fit to the proton data only due to their
identical energy dependence.}
have been fitted
to the existing experimental data for the angle-integrated cross section
for $\rho^0$
 production on hydrogen \cite{proton_rho0}.
The slope parameter $B$ in the exponential $t$-distribution was fixed at
 $B$ = 6 GeV$^{-2}$
which is known experimentally.\footnote{
The results are rather stable against
a small variation of $B$ in the range B = 6 $\pm$ 2 GeV$^{-2}$.}
In order to avoid poorly understood contributions of baryonic
resonances we have limited our fit to $W >$ 3 GeV.
The simple form of the amplitude above is obviously not adequate
for large $Q^2$ where genuine hard QCD processes take place.
Consequently we have included in our fit only experimental data points
with $Q^2 <$ 10 GeV$^2$.
In Fig.\ref{fig_map_txt}, together with all available $\rho^0$
electroproduction data, we have shown these kinematical boundaries.
The large filled circles denote the kinematical loci
where the HERMES
 analysis of charged-pion semi-inclusive data was made.
With the above cuts we get from the fit:
$C_{\Pom}$ = 1.57 $\mu b^{1/2} GeV^{-1} $,
$C_{IS}+C_{IV}$ = 6.33 $\mu b^{1/2} GeV^{-1}$
and $\mbox{\large$\epsilon$}_{\mbox{\tiny $\Pom$}}$ = 0.102.
The quality of the fit is shown in Fig.\ref{fig_sg_r0ex}.
The corresponding $\chi^2$ = 2.34.

\begin{figure}[t]
\begin{center}
\mbox{
\epsfysize 6.5cm
\epsfbox{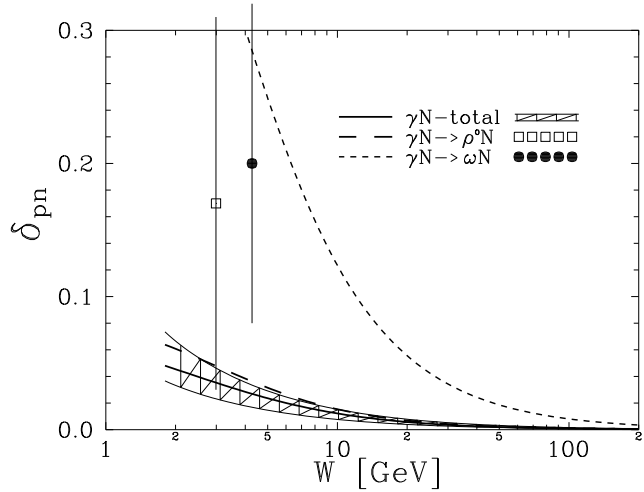}
}
\end{center}
\caption{\it
The asymmetry $\delta_{pn}$ as a function of center-of-mass energy.
The thick solid line is
obtained from the Regge parametrization adjusted to reproduce
the empirical fit for total photoproduction
(\ref{empirical_fit}) represented by the hatched band.
The dashed and dotted lines correspond to a similar asymmetry for
$\rho^0$ and $\omega$ production, repectively, and were obtained from our
Regge parametrization. The experimental points are taken from
\cite{Eisenberg} and \cite{Abramson76}.}
\label{fig_apn_el_r}
\end{figure}

To separate the sum $C_{IS}+C_{IV}$ into the isoscalar and
isovector parts one needs more information.
The size of the isovector $a_2$-exchange contribution was estimated
long ago for total photoproduction cross section (see for instance
\cite{Leith78}).
It was found empirically that the total photoabsorption cross section
on the proton and neutron can be parametrized at low energy as
\begin{equation}
\sigma_{tot} = {\tilde C}_{\Pom}
             + ({\tilde C}_f \pm {\tilde C}_{a_2}) E_{\gamma, LAB}^{-1/2}
\; .
\label{empirical_fit}
\end{equation}
In our parametrization of the $ \gamma^* p \rightarrow \rho^0 p $
data the energy dependence of the pomeron- and reggeon-exchange
contributions
 is slightly different (see Eq.(\ref{Regge_amplitude})).
Extending the validity of Regge phenomenology to both real and
virtual photons we can write somewhat schematically the amplitude
\begin{equation}
A(\gamma^* N \rightarrow \gamma^* N) =
\frac{m_V^4}{(m_V^2+Q^2)^2}
\left[
\left( \frac{1}{\gamma^2_{\rho^0}} + \frac{1}{\gamma^2_\omega} \right) \Pom
+
\left( \frac{1}{\gamma^2_{\rho^0}} + \frac{1}{\gamma^2_\omega} \right) f
\pm
\frac{2}{\gamma_{\rho^0} \gamma_\omega} a_2
\right]
\label{VDM_Regge_Compton}
\end{equation}
for Compton scattering,
\newpage
\begin{equation}
A(\gamma^* N \rightarrow \rho^0 N) =
\frac{m_V^2}{m_V^2+Q^2}
\left[
 \frac{1}{\gamma_{\rho^0}} \Pom
+
 \frac{1}{\gamma_{\rho^0}} f
\pm
 \frac{1}{\gamma_\omega} a_2
\right]
\label{VDM_Regge_rho0}
\end{equation}
for exclusive $\rho^0$ photoproduction and
\begin{equation}
A(\gamma^* N \rightarrow \omega N) =
\frac{m_V^2}{m_V^2+Q^2}
\left[
 \frac{1}{\gamma_{\omega}} \Pom
+
 \frac{1}{\gamma_{\omega}} f
\pm
 \frac{1}{\gamma_{\rho^0}} a_2
\right]
\label{VDM_Regge_omega}
\end{equation}
for exclusive $\omega$ photoproduction.
We have used the values of $\gamma_{\rho^0}$ and $\gamma_{\omega}$ from
\cite{Ioffe_book}
and put $m_{\rho} = m_{\omega} \equiv m_V$.
Please note that $\Pom$, $f$ and
$a_2$ corresponding to the reggeon-exchange amplitudes on the hadronic
level are the same in all these reactions.
Different factors in front of these hadronic amplitudes give different
strength of each contribution in different reactions.
We have adjusted the relative strength of the $a_2$-contribution
compared to the $f$-contribution
in the Compton scattering amplitude (\ref{VDM_Regge_Compton})
to reproduce the empirical low-energy parametrization
(\ref{empirical_fit}) for
$\sigma_{tot}(\gamma p)$ and $\sigma_{tot}(\gamma n)$.
In Fig.\ref{fig_apn_el_r} we compare
$\delta_{pn}^{Compton} =
\frac{\sigma_{tot}(\gamma p) - \sigma_{tot}(\gamma n)}
      {\sigma_{tot}(\gamma p) + \sigma_{tot}(\gamma n)}$
given by the obtained parametrization
with that from the empirical fit.
Shown is the band due to the uncertainties of parameters
from \cite{Leith78} and the best representation of the empirical formula
by our Regge parametrization (\ref{Regge_amplitude}).
Although we reproduce the empirical fit rather well,
in our case there is a different relative strength of the $f$ and $a_2$
contributions to the Compton amplitude.
This difference is caused by the energy dependence
of the pomeron exchange as opposed to the constant assumed
in the empirical fit \cite{Leith78}.

Having fixed parameters in (\ref{VDM_Regge_Compton})
we can calculate the corresponding proton-neutron asymmetry
for $\rho^0$ and $\omega$ production:
$\delta_{pn}^{\rho^0}$, $\delta_{pn}^{\omega}$,
which are also shown in Fig.\ref{fig_apn_el_r}.
We have shown in addition experimental results for
$\rho^0$ \cite{Eisenberg} and $\omega$ \cite{Abramson76} photoproduction
which are consistent with our parametrization.
While the asymmetry of the cross sections for the $\rho^0$
production is similar to the Compton case, the asymmetry for
the $\omega$ production is considerably larger.\footnote{
We have neglected here the pion-exchange mechanism which
would decrease $\delta_{pn}^{\omega}$ at energies $W <$ 5 GeV.}
The latter may also be important in the context of $\bar d - \bar
u$ asymmetry. However, the absolute normalization of the corresponding
cross section for the $\gamma^* N \rightarrow \omega N$ reaction
is not well known, at least in the region of
the HERMES kinematics. It is expected to be considerably
smaller than for the $\rho^0$ production.

Although the difference of the cross sections for exclusive $\rho^0$
production on the proton and neutron targets is small, the effect
of this mechanism on the $\bar d - \bar u$ asymmetry is not negligible
at all.
\begin{figure}[t]
\begin{center}
\mbox{
\epsfysize 7.2cm
\epsfbox{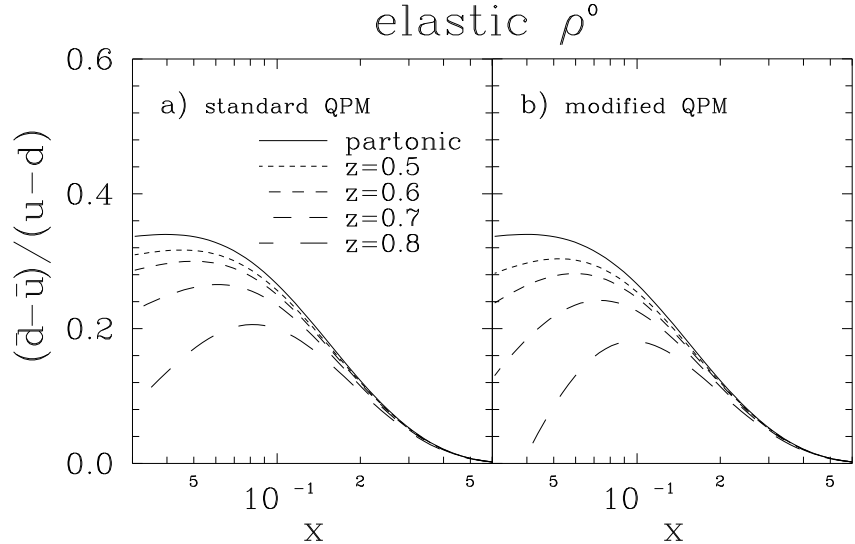}
}
\end{center}
\caption{\it
The true (solid) and the modified by the exclusive $\rho^0$
production $(\bar d - \bar u) / (u - d)$ as a function of
Bjorken-$x$ for different $z$ and typical HERMES $W$ = 5 GeV.
As in the central VDM case,
in panel (a) $Q_0^2$ = 0 (standard partonic component) and in panel
(b) $Q_0^2$ = 0.8 GeV\,$^2$ (rescaled partonic component).}
\label{fig_hermes_e}
\end{figure}
In Fig.\ref{fig_hermes_e}, we show the corresponding modification of
the quantity $\frac{\bar d - \bar u}{u-d}$ in the same way as before
for the central VDM contribution.
This modification may even be underestimated, as it is based on the
Regge-inspired parametrization of the cross section for the
$\gamma^* n \rightarrow \rho^0 n$
reaction which overestimates the $\rho^0$ production on nuclei.
This can be seen in Fig.\ref{fig_sg_r0ex} where the NMC experimental
points lie below the Regge-parametrization.

\subsubsection{QCD approach}

For sufficiently large $Q^2$ \footnote{It is not clear at present
how large the virtuality should actually be for the applicability
of this formalism.} the elastic vector meson production
can be calculated perturbatively in the formalism of off-forward
parton distributions (OFPD's) \cite{VGG98,MPW98}.
It was argued that
 only the cross section for longitudinally polarized
photons,
 where end-point contributions are suppressed,
can be calculated reliably \cite{CFS97}.
The cross section for the exclusive reaction
$\gamma_L^* + N \rightarrow V_L + N$ can be calculated in the standard
way as
\begin{equation}
\frac{d \sigma_{LL}^{\gamma + N \rightarrow V + N}}{dt} =
\frac{1}{16 \pi s^2} \frac{1}{2} \sum_{\lambda_N \lambda_N'}
| {\cal M}_{\lambda_N, \lambda_N'}^{0, 0}(t) |^2
\; ,
\label{t_dependence}
\end{equation}
where $\lambda_N$ and $\lambda_N'$ are helicities of the incoming and
outgoing nucleons, respectively. The two zeros in the upper index row
of the matrix element correspond to longitudinal photons and helicity
0 of the produced vector meson.
The amplitude of the two-quark exchange mechanism for vector meson
production was calculated for the first time in \cite{VGG98,MPW98}.
The total longitudinal cross section can be obtained by integrating
(\ref{t_dependence}) over $t$ in the kinematically allowed interval.

In the formalism first proposed
 by Ji \cite{Ji97}, neglecting
transverse momenta of quarks in the nucleons
 and in the vector meson,
the leading order amplitude reads \cite{VGG99}
\begin{eqnarray}
{\cal M}_{\lambda_N, \lambda_N'}^{0, 0}(t) =
- i \frac{4}{9} \frac{1}{Q}
\int_{0}^{1} dz
 \frac{\Phi_V(z)}{z(1-z)} \;
\frac{1}{2}
\int_{-1}^{1} dx
 \left[
\frac{1}{x - \xi + i \epsilon} + \frac{1}{x + \xi - i \epsilon}
\right]      \cdot
\nonumber \\
(4 \pi \alpha_s) \; H_N^V(x,\xi,t) \;
 \bar N(p',\lambda_N') \gamma \cdot n N(p,\lambda_N) \; ,
\label{qe_amplitude}
\end{eqnarray}
where $\Phi_V(z)$ is the distribution amplitude and $H_N^V(x,\xi,t)$
is a generalized function related to so-called skewed quark distributions
in the nucleon.
For the electroproduction of $\rho^0$ mesons we are interested in here
one gets
\begin{equation}
H_{N}^{\rho_0}(x,\xi,t) = \frac{1}{\sqrt{2}}
\left[
\frac{2}{3} H^{u/N}(x,\xi,t) + \frac{1}{3} H^{d/N}(x,\xi,t)
\right] \: .
\end{equation}
The functions $H^{u/N}(x,\xi,t) \equiv u_N(x,\xi,t)$ and
$H^{d/N}(x,\xi,t) \equiv d_N(x,\xi,t)$ are the non-diagonal, off-forward
quark distributions. In this subsection we shall concentrate
on the relative magnitude of
the cross sections for $\rho^0$ production off the
neutron and proton. Therefore the approximation relying on
the replacement of $\xi \rightarrow x$ (i.e.
using familiar diagonal quark distributions) seems sufficient for
our present purpose. Consequently we shall take
\begin{eqnarray}
H^{u/N}(x,\xi,t) &=& u_N(x) \cdot D(t) \nonumber \\
H^{d/N}(x,\xi,t) &=& d_N(x) \cdot D(t)
\label{quarks}
\end{eqnarray}
for $x > 0$ and
\begin{eqnarray}
H^{u/N}(x,\xi,t) &=& -\bar u_N(x) \cdot D(t)  \nonumber \\
H^{d/N}(x,\xi,t) &=& -\bar d_N(x) \cdot D(t)
\label{antiquarks}
\end{eqnarray}
for $x < 0$. The above Ansatz assumes factorization of $x$ and $t$
dependences.
Thus the whole $t$-dependence will be contained in one universal
function $D(t)$, common for all flavours.
We shall try exponential and dipole form factors which provide
a good representation of experimental data for exclusive
$\rho^0$ production. The factorized form has the advantage that the total
longitudinal cross section can be obtained analytically.

In the present analysis we have neglected the tensor magnetic-type
E-terms (see \cite{VGG99})
\footnote{As far as we know these terms were never estimated
in the literature.}
which may be expected to be important only at large $t$ and lead therefore
to a rather small contribution to the total cross section.

The integral in the amplitude given in (\ref{qe_amplitude}) can be
calculated in the standard way by splitting the integral into
a real principle value and an imaginary $\delta$-function.

\begin{figure}[t]
\begin{center}
\mbox{
\epsfysize 7.0cm
\epsfbox{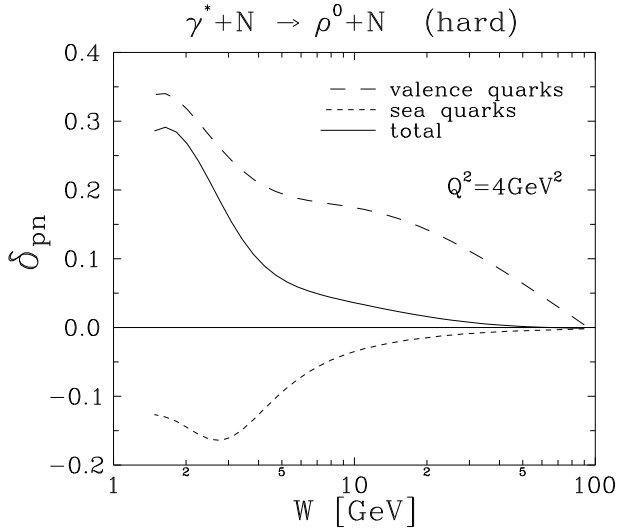}
}
\end{center}
\caption{\it
$\delta_{pn}^L$ for quark-exchange mechanism as a function of
center-of-mass energy. The solid line includes both valence
and sea quarks. For completeness we show also
the asymmetry for valence (dashed line) and sea (dotted)
quarks exclusively.
}
\label{fig_apn_el_h}
\end{figure}

The cross section asymmetry defined as
\begin{equation}
\delta_{pn}^L \equiv \frac{
\sigma_L^{\gamma^* p \rightarrow \rho^0 p} -
\sigma_L^{\gamma^* n \rightarrow \rho^0 n}    }
{
\sigma_L^{\gamma^* p \rightarrow \rho^0 p} +
\sigma_L^{\gamma^* n \rightarrow \rho^0 n}    }
\label{ntop_ratio}
\end{equation}
calculated according to (\ref{t_dependence}) and (\ref{qe_amplitude})
is shown in Fig.\ref{fig_apn_el_h} as a function of photon-nucleon
center-of-mass
 energy $W$ for $Q^2$ = 4 GeV$^2$, typical for
the HERMES kinematics.
As in previous calculations the quark distributions in
Eqs.(\ref{quarks}) and
 (\ref{antiquarks}) were taken from \cite{GRV94}.
If only valence quark distributions are taken into account
there is a relatively large asymmetry between the scattering off
the proton and neutron targets. The inclusion of sea quarks
decreases the asymmetry, which vanishes completely at large energy
(small $x$ for a fixed $Q^2$). At the energy of the HERMES experiment
$W \sim$ 5 GeV there is about 10 percent asymmetry i.e.\  about a factor
2 more than in the Regge approach.

\begin{figure}
\mbox{
\epsfysize 6.0cm
\epsfbox{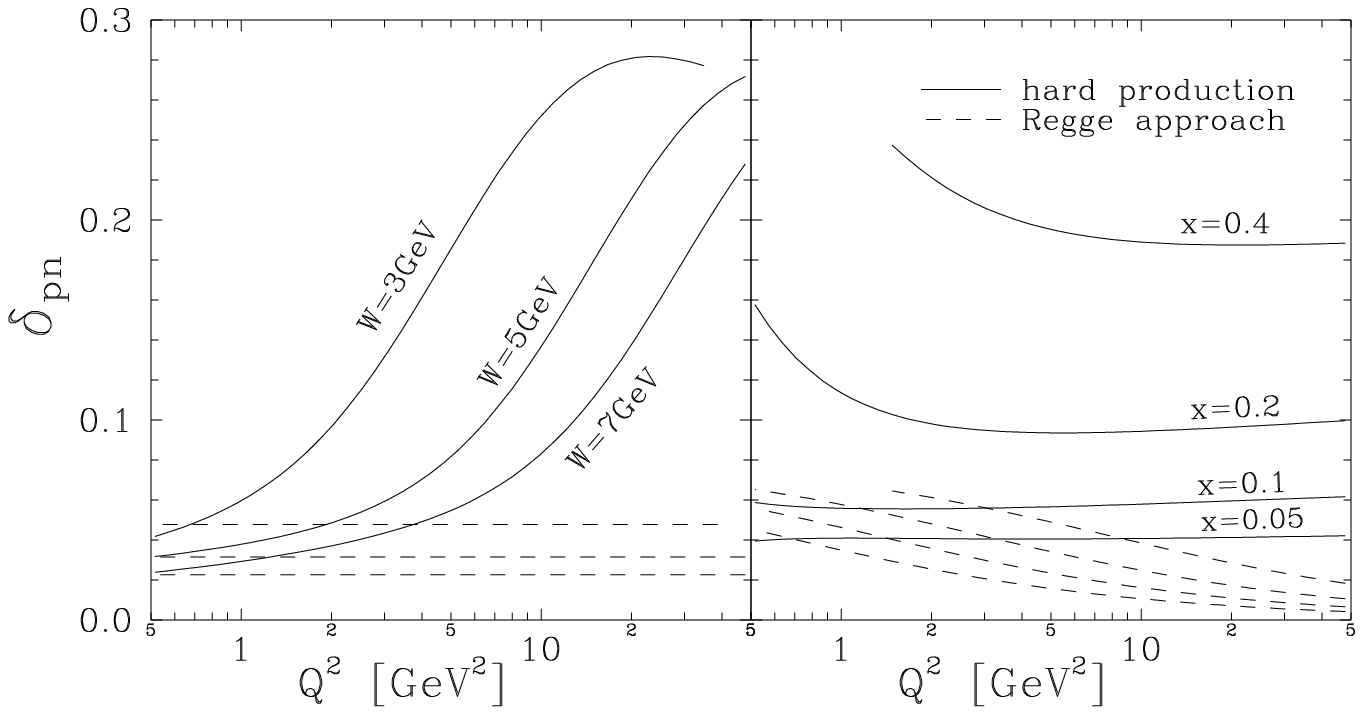} }
\caption{\it
$\delta_{pn}^L$ as a function of $Q^2$ in the Regge (dashed)
and QCD inspired (solid) approaches for:
 different fixed $W$ = 3, 5, 7 GeV (left panel) and
 different fixed Bjorken-$x$ = 0.05, 0.1, 0.2, 0.4 (right panel).
}
\label{fig_asy_wx}
\end{figure}
In Fig.\ref{fig_asy_wx} we compare the asymmetry obtained within
the Regge approach
discussed in the previous section and within the perturbative formalism
discussed here for fixed $W$ (left panel) and fixed $x$ (right panel)
as a function of $Q^2$.
There is a substantial difference between the Regge approach, where
$W$ is the variable relevant for the proton and neutron asymmetry,
and the QCD approach,
where it is rather Bjorken-$x$ which controls the asymmetry.
In the Regge approach, if $W$ is fixed the asymmetry is
practically independent of $Q^2$ and varies strongly for fixed
Bjorken-$x$.
In contrast, in the perturbative approach for
fixed Bjorken-$x$ the asymmetry only weakly depends on
the photon virtuality, as can be seen from Fig.\ref{fig_asy_wx}b.
If the center-of-mass energy is fixed instead, the $Q^2$ dependence of
the ratio is much stronger.
As can be seen from Fig.\ref{fig_asy_wx} the asymmetry between proton
and neutron target becomes larger for larger photon virtuality.

We have calculated only longitudinal cross sections which dominate
at large $Q^2$. At small $Q^2$ the transverse cross section
becomes equally important.
Although it is not possible to make a rigorous calculation
for the transverse cross sections, it is natural to expect
$\delta_{pn}^T$ and $\delta_{pn}^L$ to be similar.
Therefore we expect that the asymmetry for longitudinal
cross sections should be a reasonable estimate of
the asymmetry of the total (longitudinal+transverse) cross sections.

The description of the experimental data for exclusive
$\rho^0$ meson production by means of the hard mechanism
(not discussed here) is not as good as by means of the Regge approach.
The absolute normalization of the cross section depends
on transverse momentum distributions of quarks in the nucleon
and in the produced $\rho$ meson \cite{VGG99} which are not
fully understood at present.
Therefore we shall not calculate here the corresponding modification of
the measured $(\bar d - \bar u) / (u - d)$. Such an analysis
requires first of all a good description of the absolute
value of the cross section.
We expect, however, at least as big a modification
as in the Regge case.

\subsubsection{Charged $\rho$ mesons}

Above we have considered only neutral $\rho$ mesons.
Charged $\rho$ mesons can also be a source of charged pions due to
their decay mode
$\rho^{\pm} \rightarrow \pi^{\pm} \pi^0$. Experimentally
the cross section for exclusive charged $\rho$ mesons is much less well
known than that for the neutral $\rho$ mesons. It can be estimated
within the QCD formalism of the OFPDs approach as that sketched
for neutral $\rho$ mesons using symmetry relations for
the matrix elements \cite{MPW99}.
Because the cross section for exclusively produced charged mesons
depends on quark distributions in the nucleon differently than
in the QPM formula (\ref{semi_parton})
the contribution of charged $\rho$ mesons will certainly
modify the $\bar d - \bar u$ extracted by means of
(\ref{exp_extract}).
Because the result depends on
the magnitude of the charged-meson production which is rather difficult
to predict in the QCD-type calculations (off-diagonal effects,
the choice of the scale of the running coupling constant,
inclusion of transverse momenta)
we shall leave the problem for a separate more refined analysis.

\section{Comments on nuclear effects in the deuteron}

So far we have followed the HERMES collaboration
and neglected all nuclear effects in the deuteron i.e.
assumed that
\begin{equation}
\sigma(\gamma^* d \rightarrow \pi^{\pm}) =
\sigma(\gamma^* p \rightarrow \pi^{\pm}) +
\sigma(\gamma^* n \rightarrow \pi^{\pm}) \; .
\end{equation}
The theory of nuclear effects in semi-inclusive processes is less
developed than in the inclusive case. Let us consider a simple example of
an $x$-independent
 relative nuclear effect of size $\kappa$,
universal for produced $\pi^+$ and $\pi^-$,
\begin{equation}
\sigma(\gamma^* d \rightarrow \pi^{\pm}) =
(1 - \kappa) \left( \sigma(\gamma^* p \rightarrow \pi^{\pm}) +
 \sigma(\gamma^* n \rightarrow \pi^{\pm}) \right) \; .
\label{nucl_effect}
\end{equation}
Then the semi-inclusive cross sections for pion production on
the neutron extracted from the deuteron target data are
\begin{eqnarray}
\sigma_{meas}(\gamma^* n \rightarrow \pi^{\pm}) &=&
\sigma(\gamma^* d \rightarrow \pi^{\pm}) -
\sigma(\gamma^* p \rightarrow \pi^{\pm}) \nonumber \\
 &=&
(1 - \kappa) \sigma(\gamma^* n \rightarrow \pi^{\pm}) -
\kappa \sigma(\gamma^* p \rightarrow \pi^{\pm}) \\
 &=&
\sigma(\gamma^* n \rightarrow \pi^{\pm}) - \kappa
\left( \sigma(\gamma^* p \rightarrow \pi^{\pm}) +
       \sigma(\gamma^* n \rightarrow \pi^{\pm}) \right)
 \nonumber
\end{eqnarray}
i.e.\ biased by the assumed nuclear effect $\kappa$
in the deuteron.

Thus the differences
$\sigma(\gamma^* p \rightarrow \pi^{\pm}) -
 \sigma(\gamma^* n \rightarrow \pi^{\pm})$
needed in Eq.(\ref{exp_extract}) are replaced by
\begin{eqnarray}
& \sigma(\gamma^* p \rightarrow \pi^{\pm}) -
\sigma_{meas}(\gamma^* n \rightarrow \pi^{\pm}) = & \nonumber \\
& = (1 + \kappa) \sigma(\gamma^* p \rightarrow \pi^{\pm}) -
(1 - \kappa) \sigma(\gamma^* n \rightarrow \pi^{\pm}) & \\
& = \sigma(\gamma^* p \rightarrow \pi^{\pm}) -
    \sigma(\gamma^* n \rightarrow \pi^{\pm}) +
\kappa \left(
 \sigma(\gamma^* p \rightarrow \pi^{\pm}) +
 \sigma(\gamma^* n \rightarrow \pi^{\pm})
 \right)
\nonumber
\end{eqnarray}
%

\begin{figure}[t]
\begin{center}
\mbox{
\epsfysize 7.8cm
\epsfbox{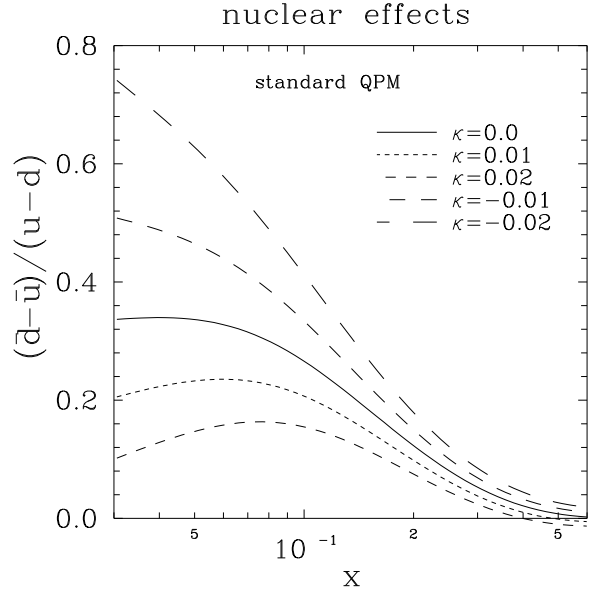}
}
\end{center}
\caption{\it
The true (solid) and the modified by the nuclear effects
$(\bar d - \bar u) / ( u - d )$ as a function
of Bjorken-$x$ for $W$ = 5 GeV and different values of $\kappa$.
}
\label{fig_hermes_s}
\end{figure}
In Fig.\ref{fig_hermes_s} we show the nuclear effects on
$ \frac{\bar d - \bar u}{u - d} $
for $\kappa$ = 0.02, 0.01, 0.0, -0.01, -0.02,
i.e.\ in the range
 known from inclusive DIS.
These effects are independent
 of $z$ by assumption (\ref{nucl_effect}).
Following the inclusive case, for small values of Bjorken $x < 0.1$ a
shadowing,
 i.e.\ $\kappa > 0$ is expected
which means that the asymmetry obtained when neglecting
nuclear effects is underestimated (see Fig.\ref{fig_hermes_s}).

The shadowing
 leads to an effect opposite to that for the resolved photon
component discussed earlier in this paper.
For somewhat larger $x$ an antishadowing due to excess pions is
not excluded. For still larger $x$ a nuclear binding and Fermi motion
corrections come into play.

Summarizing, we have shown that even small nuclear effects,
of the order of just a few percent, lead to considerable consequences
for
 the $\bar d - \bar u$ asymmetry.
 Nuclear effects are expected
to be $x$ and $Q^2$ dependent.
In the present analysis we have shown only a band of
uncertainties due to nuclear effects.
A more precise determination of the $x$ or $Q^2$ dependence
requires a more microscopic calculation which goes beyond the scope of
the present paper. This is, however, necessary if the $\bar d - \bar u$
asymmetry is to be extracted from semi-inclusive data.

\section{Consequences for the HERMES experiment}

Having discussed each of the nonpartonic effects separately we will
now attempt to combine them and try to understand their net effect
on the measured $\bar d - \bar u$ asymmetry, and the consequences
of this for the HERMES data \cite{HERMES_dbar_ubar}.

Before a numerical estimation of the total nonpartonic effect
we would like to discuss briefly a subtle problem.
At first sight it seems that fragmentation functions fitted to
experimental data effectively contain nonpartonic effects.
This is most probably not true as nonpartonic components
are higher-twist effects i.e.\  are strongly $Q^2$-dependent
in contrast to the leading twist.

In section \ref{choice_ff} we have selected the fragmentation functions
which describe the pionic yields well.
There is a danger a priori that explicit inclusion of the nonpartonic
effects discussed in the present paper  may worsen the description
of pionic spectra.
\begin{figure}[t]
\begin{center}
\mbox{
\epsfysize 9.0cm
\epsfbox{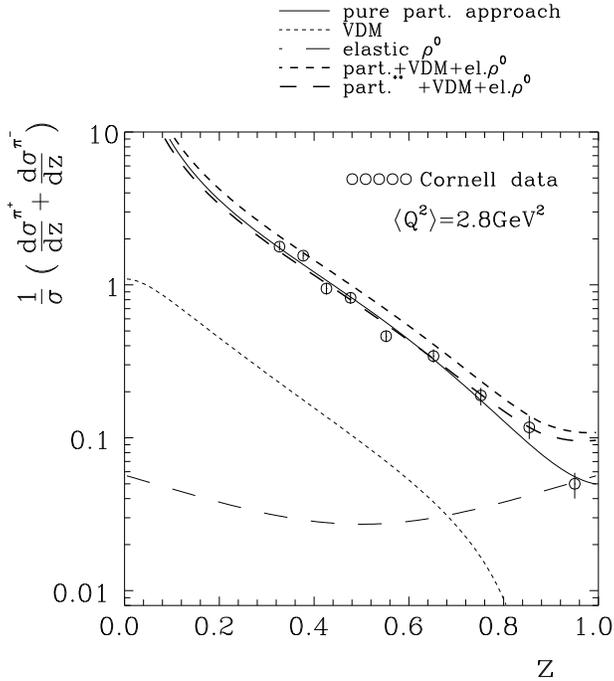}
}
\end{center}
\caption{\it
Multiplicity distribution of the charged pions.
Contributions of different mechanisms are shown separately.
The total effect calculated in two different ways discussed
in the text is also presented.
}
\label{fig_mult_all}
\end{figure}
In Fig.\ref{fig_mult_all} we show the multiplicity
of charged pions as a function of $z$ for $<\!Q^2\!>$ = 2.8 GeV$^2$
(compare with the left panel of Fig.\ref{fig_cha_pi}).
Together with the corresponding Cornell data \cite{Drews78} we show
the contributions due to different mechanisms separately.
The contributions of nonpartonic mechanisms are considerably smaller
than the main partonic contribution, however not negligible.
If we add all of them together we obtain the multiplicity (thick
short-dashed line) over the experimental points.
Another way to incorporate the partonic and nonpartonic
components was proposed in Ref.\cite{SU_inclusive} for the inclusive case.
As can be seen from Fig.\ref{fig_mult_all}, extension of this model
to the semi-inclusive case, i.e.\  rescaling of the partonic component
as in Eq.(\ref{SU_semi}), (thick long-dashed line) provides
a very good description of the experimental multiplicities.
This approach treats all contributions explicitly, which seems
more consistent than the approach mentioned above including them
all effectively into fragmentation functions.
It is also consistent with the inclusive structure function model
\cite{SU_inclusive}.

\begin{figure}
\begin{center}
\mbox{
\epsfysize 12.0cm
\epsfbox{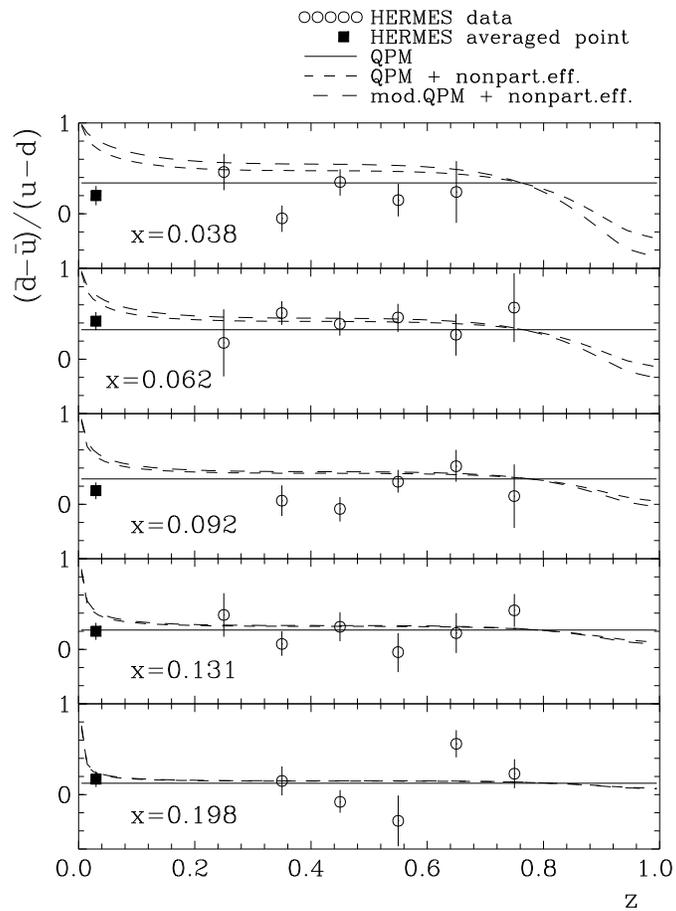}
}
\end{center}
\caption{\it
$\frac{\bar d - \bar u}{u - d}$ as a function of $z$ for
different bins of $x$. The experimental data are from
\cite{HERMES_dbar_ubar}.
}
\label{fig_hermes_z}
\end{figure}
In Fig.\ref{fig_hermes_z} we show $\frac{\bar d - \bar u}{u - d}$
as a function
 of $z$ for different bins of Bjorken-$x$
(the averaged value of
 $x$ is given in the figure).
The data points are taken from \cite{HERMES_dbar_ubar}.
The filled squares correspond to the data averaged over $z$.
The QPM ($z$-independent) prediction with leading order quark distributions
from \cite{GRV94} is shown for reference by the solid line.
The results of the calculation including
 partonic, VDM and
elastic $\rho^0$ contributions are given by the dashed
 line
(with the QPM contribution calculated in the standard way) and
 long-dashed line
(with the QPM contribution modified as in Eq.(\ref{SU_semi})).
As in the previous calculations the Field-Feynman fragmentation
functions \cite{FF77} were used here.

As can be seen from the figure there is a significant deviation of the
``measured'' asymmetry from the ``real'' one (the difference between
the dashed
 and solid lines), especially for small values of
Bjorken-$x$. The effect is bigger for the
``modified-QPM'' approach i.e.\ for the more consistent one.
The nuclear effects most probably will introduce further deviation which
is, however, difficult to estimate numerically.
A significant $z$-dependence casts doubts on the averaging
in $z$ at least in the whole range from 0 to 1. However, as seen from
the figure the experimental statistics and the $z$-range of the HERMES
experiment do not allow this dependence to be identified.

It is worth noting that
the modification of the QPM result depends strongly on
the fragmentation function used to calculate the ratio $D_{-}/D_{+}$.
The Field-Feynman fragmentation functions used
in the present analysis provide a good representation of the data
from both EMC and HERMES (preliminary) (see Fig.\ref{fig_unf_fa_r})
i.e.\ are close to those used in extraction of the $\bar d - \bar u$
asymmetry by the HERMES collaboration.
One should also remember that the effect of elastic $\rho^0$ production
is model dependent, being generally larger in the hard production mechanism
than in the Regge model.
If we take our Regge result at face value we argue that
the $\bar d - \bar u$
 asymmetry extracted by the HERMES collaboration
is rather overestimated.
In the present analysis we do not attempt to correct the HERMES data
for the effects discussed here. This requires a separate analysis
including efficiencies of the HERMES apparatus as well as
knowledge of the many cuts
 used
 in their analysis.

\section{Conclusions}

Extraction of parton densities is one of the main goals
of high-energy physics. It was proposed some time ago
how to use semi-inclusive production of charged pions
to determine both unpolarized and polarized parton densities
in the nucleons. Recently this idea was put into practice
in both cases. Such analyses assume implicitly
the validity of the quark-parton model. In a recent
work of two of us \cite{SU_part_viol} we have shown a breaking
of the parton model in inclusive DIS at photon virtuality $Q^2$
as large as 5-7 GeV$^2$ which is bigger than commonly perceived.
The modern experiments analyzing semi-inclusive production of pions
are performed in a similar range of $Q^2$. In general
the situation in semi-inclusive reactions can be even more complex and
subtle. In the present analysis we have made a first attempt to
determine the nonleading mechanisms.

We have estimated a few
 effects beyond the quark-parton model which
may influence
 the extraction of the $\bar d - \bar u$ asymmetry from
semi-inclusive production of pions in DIS.

Based on the analysis of hadronic data we have found that
the interaction of the resolved hadron-like photon
with the nucleon may lead to an artificial enhancement of
the measured $\bar d - \bar u$ asymmetry in the region of small
 $x$.
Next, we have investigated the elastic production of
$\rho^0$ mesons by a virtual photon on the proton and neutron targets
based on two
 different models. Unequal cross sections for
proton and neutron targets also lead to an artificial
modification of the $\bar d - \bar u$ asymmetry extracted based
on QPM formulae. The effect found is opposite to the
effect due to the resolved photon component.
These two effects cancel only in a narrow range of $z$.
The net effect turned out to be $z$-dependent
invalidating somewhat averaging in $z$ as done recently
in \cite{HERMES_dbar_ubar}.

We suggest that instead of averaging over a broad range of $z$ one
could try to select the region of $z$ ($x$- and
$Q^2$-dependent) where the influence
of nonpartonic effects is small. Unfortunately this can only be done
at the expense of lowering the statistics considerably.
An optimal choice of kinematical cuts in $x$, $Q^2$
 and $z$
requires a more detailed study. Clearly, increasing of $Q^2$ looks
helpful. This could be realized by HERMES or by an as yet unproposed
experiment for COMPASS at CERN.

Nuclear effects, even small ones, may also cloud the extraction
of the true $\bar d - \bar u$ asymmetry. To our knowledge
there is no reliable estimate of such effects for semi-inclusive
production of pions.

Although in the light of the present analysis
the precise direct extraction of $\bar d - \bar u$ is rather
difficult, the semi-inclusive data can be used for tests of parton
distributions,
in particular the difference between $\bar d$ and $\bar u$, provided
nonpartonic and nuclear effects are understood and included in the analysis.

Finally we would like to point out that some of the effects discussed
in the present paper may also influence the extraction of the polarized
quark distributions from semi-inclusive production of pions in DIS.
This will be a subject of a separate analysis.

\section{Appendices}
\appendix

\section{\large \bf 
Parametrizing the $\pi p \rightarrow \pi X$ spectra in the pion hemisphere} 
\label{app_bosetti} 

The mechanism of inclusive pion production in hadronic reactions
is in general not well understood. It is believed that rather
soft processes dominate. Some progress was made recently in studying
inclusive pion production in polarized scattering (for a review see
\cite{LB00}). In proton-proton scattering the large-$x_F$ region depends
on the flavour structure of the outgoing pion. This part of the spectrum
was explained e.g.\  in the recombination model \cite{DH77}, fusion model
\cite{Liang-Meng} and recently in the meson cloud model
\cite{NSSS99,CDNN99,Boros99}.
The central (mid-rapidity) part of the spectrum seems to be flavour
independent \cite{Liang-Meng}.

The inclusive spectra of pions in pion-nucleon scattering are,
in general, even less well understood.
For beam-like pions the large-$x_F$ part of the spectra
seems to be dominated by diffractive processes due to pomeron or reggeon
($f$ or $\rho^0$) exchanges. For both beam-like and beam-unlike spectra
one may expect a nonnegligible contribution from the decay of $\rho$ mesons
produced in peripheral processes $\pi + p \rightarrow \rho + X \rightarrow
(\pi+\pi) + X$, dominated by pion exchange.

The most complete experimental data for the
 $\pi^{\pm} p \rightarrow \pi^{\pm} X$ reactions
were collected by the ABBCCHW
collaboration at the CERN hydrogen bubble chamber \cite{Bosetti73}.
A detailed analysis of the $\pi^+ p \rightarrow \pi^{\pm} X$
and $\pi^- p \rightarrow \pi^{\pm} X$ two-dimensional spectra
\cite{Bosetti73}
combined with a general understanding of the reaction mechanism
have shown that the spectra of all 4 reactions (4 $\times$ 10 = 40 spectra)
can be represented by the following six-component Ansatz:
\begin{eqnarray}
\frac{d \sigma^{i \rightarrow j} } { dx_F dp_{\perp}^2 }  &=&
C_{soft} \left(1 - \frac{\eta}{\eta_{max}} \right)^{p_{soft}}
\frac{\partial \eta}{\partial x_F}
\cdot
e^{-B_{soft} p_{\perp}^2 }
\nonumber \\
&+&
C_{hard}^{ij} \cdot f_{hard}(x_F)
\cdot
e^{-B_{hard} p_{\perp}^2 }
\nonumber \\
&+&
C_{cen}^{ij} \cdot e^{ \frac{x_F^2}{2 \sigma_{cen}^2} }
\cdot
e^{-B_{cen} p_{\perp}^2 }
\nonumber \\
&+&
C_{\Pom} \cdot x_F (1-x_F)^{\alpha_{\Pom}}
\cdot
e^{-B_{\Pom} p_{\perp}^2 }
\nonumber \\
&+&
C_R \cdot x_F (1-x_F)^{\alpha_R}
\cdot
e^{-B_R p_{\perp}^2 }
\nonumber \\
&+&
C_{\rho}^{ij} \cdot f_{\rho}(x_F)
\cdot
e^{-B_{\rho} p_{\perp}^2 } \; ,
\label{pip_pi_parametrization}
\end{eqnarray}
where the maximal rapidity $\eta_{max} = \eta_{max}(p_{\perp}^2)$.

Each of the components above corresponds to a distinct physical
mechanism, the first three to central processes and the last
three to peripheral processes.
By analogy with $pp$ collisions \cite{Liang-Meng} we have assumed
one universal (flavour independent) soft component
and allow for different normalization of flavour dependent components,
called here hard due to their transverse momentum dependence.
We have found $B_{soft}$ = 8.5 GeV$^{-2}$ and
$B_{hard}$ = 3.0 GeV$^{-2}$), consistent with characteristic slopes
of soft and hard processes.
The parametrization of the soft component
gives a multiplicity rising with the energy in the entrance channel.
Some models in the literature predict a growth of the flavour asymmetric
part of the inclusive cross section with energy, some predict that it
should stay almost constant.
Therefore in the present paper we have tried different functional forms for
the phenomenological flavour dependent part called here the hard component:
\begin{equation}
f_{hard}(x_F) =
\begin{cases}
(1-x_F^2)^{p_{hard}} \, , \\
(1-\frac{\eta}{\eta_{max}})^{p_{hard}}
\frac{\partial \eta}{\partial x_F} \, .
\end{cases}
\label{hard_longitudinal}
\end{equation}

In the present schematic parametrization we have taken
$\alpha_{\Pom}$ = 1 and $\alpha_R$ = 0.5, i.e.\  we have neglected
$t$-dependence of the Regge trajectories. By comparison with
the experimental data we found $B$ = 5.0 GeV$^{-2}$.
The extra factor $x_F$ in front of the diffractive components was added
to extrapolate smoothly down to $x_F=0$.
\begin{figure}[t]
\begin{center}
\mbox{
\epsfysize 8.5cm
\epsfbox{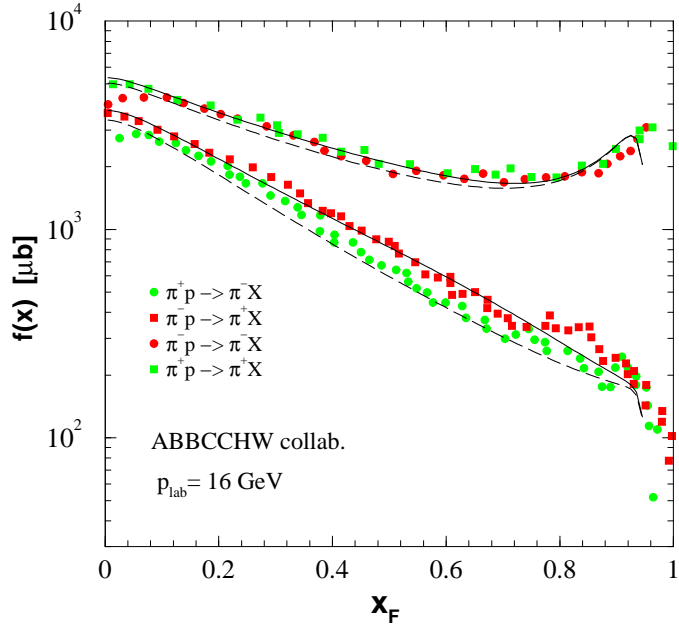}
}
\end{center}
\caption{\it
An example of the quality of the 
parametrization (\ref{pip_pi_parametrization}).
The experimental data points were scanned from Fig.1 in \cite{Bosetti73}.
}
\label{fig_bos_it_r}
\end{figure}
We have found that peripheral production of $\rho$ mesons due to pion
exchange and their subsequent decay constitutes a nonnegligible
source of charged pions. Inspired by the pion exchange model and
consistently with the data we have fixed the relations between
normalization constants for such different processes
\begin{displaymath}
C_{\rho}^{++} \approx C_{\rho}^{--}
 \approx 2 C_{\rho}^{-+} \approx 2 C_{\rho}^{+-}.
\end{displaymath}
The functional form of $f_{\rho}(x_F)$ has been taken from a
schematic model calculation and one normalization
parameter $C_{\rho}$ was fitted to the two-dimensional spectra,
which is possible
because this mechanism dominates the beam-unlike spectra at $x_F >$ 0.7.

The nature of the phenomenological very central, very soft ($B_{cen}$ =
20-30 GeV$^{-2}$) component is not clear. It was introduced only to
to describe the data. It is most probably associated with pions from
the decay of nonperipheral $\rho$ mesons. We have found empirically approximately
the same relation for normalization constants $C_{cen}^{ij}$ as
for $C_{\rho}^{ij}$.

In Fig.\ref{fig_bos_it_r} we present the quality of our fit
for transverse momentum integrated $x_F$-distributions for all
four reactions $\pi^{\pm} p \rightarrow \pi^{\pm} X$.
The results are shown
in terms of the invariant single particle structure function \cite{Bosetti73}
\begin{equation}
f(x_F) = \frac{1}{\pi} \int_0^\infty
           \frac{E^*}{p^*_{max}} \frac{d^3 \sigma}{dx_F dp_{\perp}^2}
            dp_{\perp}^2  \; .
\label{inv_spstrf}
\end{equation}
%

\section{\large \bf $\rho$ meson decay functions}
\label{app_rho0_dec}

In order to calculate the decay function in the most general case
one needs to include off-diagonal elements of the density matrix
\cite{ABMP98}. In the present paper we shall neglect the probably small
off-diagonal effects.
Then the decay function $f(z)$ depends on the helicity of the parent
$\rho^0$ meson. In the general case of broad resonance
it can be calculated as:
\begin{equation}
f_{\lambda}(z_{\pi/\rho}) = \int d m_{\rho} \rho(m_{\rho})
\int d \Omega [ f_{\lambda}(\theta,\phi)
 \delta(z(\theta,\phi)-z_{\pi/\rho}) ] \; ,
\label{D_general}
\end{equation}
where $f_{\lambda}(\theta,\phi) = |Y_{1\lambda}(\theta,\phi)|^2$
is the angular distribution of pions in the rest frame of $\rho$
and $\rho(m_{\rho})$ is the spectral density.
The momentum fraction of a pion with respect to the parent
$\rho$ meson is:
\begin{equation}
z_{\pi/\rho} = \frac{p_z^{\pi}(m_{\rho}) + p_0^{\pi}(m_{\rho})}{m_{\rho}}
\approx \frac{1}{2} (1 + \cos{\theta}) \; ,
\label{z_decay}
\end{equation}
which is independent of $m_{\rho}$.
The last relation is due to the smallness of the pion mass
and must be corrected in the case of soft pions.
In the approximation used in the present paper the two integrals
in (\ref{D_general}) factorize and one easily gets
\begin{equation}
f_{\lambda}(z) \approx
\begin{cases}
6 z (1-z) \;\;\; \text{for $\lambda = \pm 1$} \,, \\
3 (2 z - 1)^2 \;\;\; \text{for $\lambda = 0$} \,,
\end{cases}
\label{D_simplistic}
\end{equation}
where above $z$ is used instead of $z_{\pi/\rho}$ for brevity.

For the exclusive reaction $\gamma^* N \rightarrow \rho^0 N$
at high energy one has
$z \equiv z_{\pi/\gamma} \approx z_{\pi/\rho}$.
For semi-inclusive production through quark hadronization the decay
function (\ref{D_general}) must be convoluted with the fragmentation
function into the $\rho$ meson $D_{q \rightarrow \rho}$.
Below we shall consider these two distinct cases
of $\rho$ meson production.

In \underline{inclusive} unpolarized production of $\rho$ mesons
one may expect approximately an equal population of different helicities
due to the complexity of the poorly understood hadronization process.
Then the effective decay function, which is averaged over $\rho$ meson
helicities, becomes
\begin{equation}
f(z) \approx \frac{1}{3} [ 2 f_1(z) + f_0(z) ] \approx const \; .
\label{D_semi-inclusive}
\end{equation}
%

\begin{figure}
\begin{center}
\mbox{
\epsfysize 8.0cm
\epsfbox{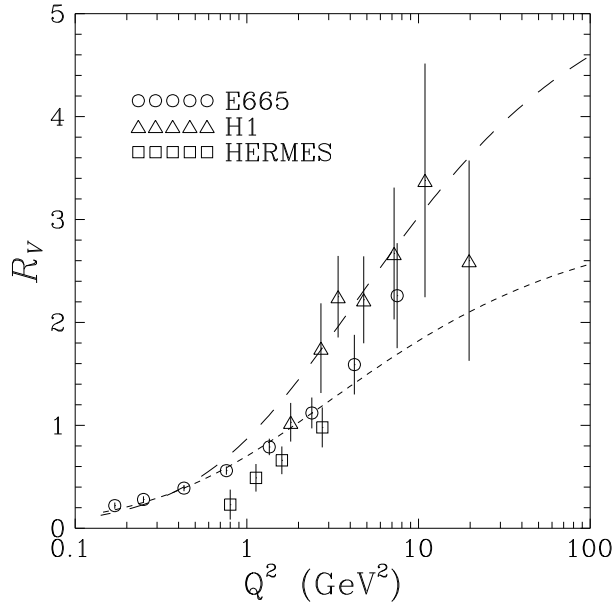}
}
\end{center}
\caption{\it
$R_V$ as a function of the photon virtuality. The experimental
data are from \cite{E665_RV,H1_RV} while the parametrizations are from
\cite{SSS99}. The long-dashed and the short-dashed lines correspond
to the 2- and 4-parameter fits, respectively.}
\label{fig_rho0_l_t}
\end{figure}
In the most general case
of \underline{exclusive} $\rho$ meson electroproduction
the angular distribution of pions
can be obtained according to the formalism of Schilling and Wolf
\cite{SW73}.
For sufficiently large energy, where $s$-channel helicity conservation
takes place, averaging over azimuthal angle, the effective decay function
can be approximated as
\begin{equation}
f(z) \approx
\frac{f_1(z) + \epsilon R_V(Q^2,W) f_0(z)}
{1 + \epsilon R_V(Q^2,W)} \; ,
\label{exc_decay_f}
\end{equation}
where the polarization parameter
$\epsilon = \epsilon(s_{e N}^{1/2}, W, Q^2) = \frac{1-y}{1-y+y^2/2}$
measures the degree of longitudinal polarization of virtual photons.
The effective decay function (\ref{exc_decay_f}) depends
on $Q^2$ due to the empirically known strong $Q^2$-dependence of
   $R_{V} = \sigma(\gamma_L N \rightarrow \rho^0 N) /
            \sigma(\gamma_T N \rightarrow \rho^0 N)$.
Some smooth dependence of $R_V$ on $W$ or $x$ is not excluded a priori.
We shall take the model parametrization of $R_V(Q^2,W)$ from
\cite{SSS99} (the 4-parameter fit)
which, as shown in Fig.\ref{fig_rho0_l_t}, adequately describes
the experimental data from \cite{E665_RV,H1_RV}.
We show there also recent HERMES experimental data \cite{HERMES_rho0}
which lie below the parametrization.
Since the HERMES data were taken on the $^3$He target
this may be partially caused by poorly understood nuclear effects.

\vskip 1cm

{\bf Acknowledgments}
We are indebted to the members of the HERMES collaboration for
discussions of their recent results and details of their apparatus
and analysis, in particular Alexander Borissov, Naomi Makins,
Pasqualle di Nezza, Valeria Muciffora and Manuella Vincter.
We are also indebted to Kolya Nikolaev for an informative discussion
and pointing us to some interesting references,
Lech Mankiewicz for a critical
 remark,
Andrzej Sandacz for pointing out some experimental data, unknown to us,
for elastic $\rho^0$ production
 and
Marc Vanderhaeghen for a discussion of the details of
their QCD-inspired calculation of the quark-exchange mechanism.
Finally we are indebted to Martin Kimber for a careful reading of the
manuscript.
This work was partially supported by the German-Polish
DLR exchange program, grant number POL-028-98.


\end{document}